\documentclass[journal]{IEEEtran}

\usepackage{amssymb,amsmath,amsfonts,bm,epsfig,graphicx,theorem,latexsym}
\usepackage{rotating,setspace,latexsym,epsf}
\usepackage{cite,authblk,epstopdf}

\newtheorem{definition}{Definition}
\newtheorem{lemma}{Lemma}

\newtheorem{remark}{Remark}

\allowdisplaybreaks[1]

\makeatletter

\begin{document}

   \author{Kaya Tutuncuoglu and Aylin Yener
   \thanks{This work was supported by NSF Grants CNS 09-64364 and CCF 14-22347. This work was presented in part at the 
	Information Theory and Applications Workshop, February 2013, 
	IEEE Information Theory Workshop, September 2013, and 
	IEEE Global Communications Conference, December 2014.}
	\thanks{The authors are with the department of Electrical Engineering, Pennsylvania State University, University Park, PA 16802 USA (e-mail: kaya@psu.edu; yener@ee.psu.edu).}
	}

   \title{Energy Harvesting Networks with Energy Cooperation: Procrastinating Policies}

\maketitle

 \begin{abstract}
 
This paper considers multiterminal networks with energy harvesting transmitter nodes that are also capable of wirelessly transferring energy to or receiving energy from other nodes in the network. In particular, the jointly optimal transmit power and energy transfer policies that maximize sum-throughput for the two-way, two-hop, and multiple access channels are identified. It is shown for nodes with infinite-sized batteries that delaying energy transfers until energy is needed immediately at the receiving node is sum-throughput optimal. Focusing on such {\em procrastinating} policies without loss of optimality, the stated joint optimization problem can be decomposed into energy transfer and consumed energy allocation problems which are solved in tandem. This decomposition is shown to hold for the finite-sized battery case as well, using partially procrastinating policies that avoid battery overflows. It is observed that for the two-hop channel, the proposed algorithm has a two fluid water-filling interpretation, and for the multiple access channel, it reduces to a single transmitter problem with aggregate energy arrivals. Numerical results demonstrate the throughput improvement with bi-directional energy cooperation over no cooperation and uni-directional cooperation.
\end{abstract}


\section{Introduction}
\label{sect_intro}

Energy harvesting wireless networks offer the possibility of perpetual network lifetime, yielding low maintenance costs and less energy storage requirements at each node \cite{kansal2007power, ulukus2015review, xiao2015survey}. However, the intermittent availability of harvested energy also renders realizing the full potential of these benefits challenging. In particular, it is possible for central nodes of the network to become energy deprived due to energy harvesting conditions, impacting the network's performance. To combat such energy outages, recent results utilizing wireless energy transfer \cite{kurs2007wireless} offer the possibility of energy cooperation for wireless ad-hoc networks in addition to signal cooperation \cite{gurakan2012energy, gurakan2012two, gurakan2013energycoop}. In this paper, we consider a generalized setting in this realm, and study multi-transmitter models where nodes can receive or transfer energy, in order to gain insights into network design by jointly optimizing data and energy transfer policies.

Optimal power allocation in energy harvesting wireless networks has recently been studied extensively. In \cite{yang2012optimal}, an energy harvesting transmitter with infinite energy storage capability is considered, and the transmission completion time minimizing power allocation in a single link is found. The short-term throughput maximization problem is studied in \cite{tutuncuoglu2012optimum}, where the setting of \cite{yang2012optimal} is extended to finite energy storage. A wireless fading channel with an energy harvesting transmitter is considered in \cite{ozel2011fading}, showing that a directional water-filling algorithm can be utilized to find the optimal power allocation. Multiterminal models, and energy harvesting transmitters and receivers are subsequently studied, see for example \cite{yang2012mac, tutuncuoglu2012sum, huang2013throughput, gunduz2011two, orhan2013throughput, ahmed2012power, luo2012optimal2, varan2013energy, tutuncuoglu2012ita, mahdavi2013energy} and references therein. In addition to optimal power policies found in  \cite{yang2012optimal, tutuncuoglu2012optimum, ozel2011fading, yang2012mac, tutuncuoglu2012sum, huang2013throughput, gunduz2011two, orhan2013throughput, ahmed2012power, luo2012optimal2, varan2013energy, tutuncuoglu2012ita, mahdavi2013energy}, alternative power management approaches such as nodes with inactive states \cite{niyato2007sleep} are also considered in previous work. 

Since energy harvesting networks may experience energy deprivation when sufficient energy is not available to harvest, they can benefit from the recent advances in wireless energy transfer. Short range energy transfer is already present in today's RFID systems \cite{want2006introduction}. Energy cooperation is a viable option in mid-range as well, with wireless energy transfer efficiency values reaching up to 40\% using coupled magnetic resonance \cite{kurs2007wireless}. This provides the possibility of energy cooperation, allowing networks to have additional control over the energy available at each node, as proposed in \cite{gurakan2012energy}. In essence, energy cooperation introduces a new dimension for network optimization in energy harvesting networks. 

The problem of optimizing energy consumption for data transmission and energy transfer is introduced in \cite{gurakan2012energy}, where a two-hop network with an energy harvesting transmitter and relay is considered, and the source can transfer energy to the relay. It is shown that throughput can be improved with respect to energy harvesting alone \cite{gunduz2011two}, even with uni-directional energy transfer. Two-way and multiple access channels with uni-directional energy cooperation are also studied by the same authors in \cite{gurakan2012two}, proposing a two-dimensional water-filling algorithm to find the optimal policy. These studies assume an infinite battery size for the transmitters. Additionally, a different line of work studies transferring energy and information jointly, see \cite{grover2010shannon, popovski2012interactive, ng2013wireless, michalopoulos2014simultaneous, michalopoulos2013relay, chen2013energy} and many others.

In this paper, we follow the model of transferring energy and data separately as in \cite{gurakan2012energy, gurakan2012two, gurakan2013energycoop}. We generalize this set up to energy harvesting nodes all of which are capable of transferring energy to one another, i.e., in any direction. 
We consider a general energy transfer model, which could be realized via various energy transfer technologies such as magnetic induction, magnetically coupled resonance, or RF harvesting. Whereas the recent paradigm of simultaneous wireless information and power transfer (SWIPT) focuses on harvesting RF signals for powering devices, and the associated trade-offs between using the energy for device operation or information decoding, we focus on optimally allocating and sharing of energy between devices over time.
As communication models, we consider those in \cite{gurakan2012energy, gurakan2012two, gurakan2013energycoop}, i.e., two-way, two-hop, and multiple access channels, allowing unrestricted energy transfers between all nodes. In addition to generalizing earlier works to unrestricted energy transfers, we also extend these models to the case where all nodes have finite battery and establish the optimal policy when the battery sizes are finite. We will see that allowing unrestricted energy transfers and limited batteries both require a careful solution methodology and bring on new design insights.

Specifically, we optimize transmit powers and energy transfers under the aforementioned general setting. For clarity of exposition, first, in Sections~\ref{sect_model}, \ref{sect_properties}, and \ref{sect_twc}, we consider the two-way channel with infinite-sized batteries at the transmitters. We prove that a subset of feasible policies, composed of those which postpone energy transfers until immediately needed, includes an optimal policy. Named {\em procrastinating policies}, this subset allows a decomposition of the joint optimization problem into separate energy transfer allocation and consumed power allocation problems. We subsequently show that the separation extends to two-hop (Section~\ref{sect_twohop}) and multiple access channels (Section~\ref{sect_mac}). We demonstrate that a generalized extension of the directional water-filling algorithm \cite{ozel2011fading} for the two-way channel solves the power allocation problem, while a single-user policy as in \cite{yang2012mac}, with scaled aggregate arrivals, suffices for the multiple access channel. In Section~\ref{sect_finite_bat}, we extend our study to transmitters with finite-sized batteries, and show that {\em partially procrastinating policies}, defined therein, are optimal. Next, we leverage a simplified version of the two-dimensional water-filling algorithm in \cite{gurakan2012energy} to solve the throughput optimization problem with joint energy transfer and transmit powers. We present numerical results in Section~\ref{sect_numerical}, demonstrating the advantage of bi-directional energy cooperation in energy harvesting networks over no cooperation \cite{gunduz2011two, yang2012optimal} and uni-directional cooperation \cite{gurakan2012energy, gurakan2012two, gurakan2013energycoop}. Section~\ref{sect_conclusion} concludes the paper.

\section{The Energy Harvesting and Energy Cooperating Two-way Channel (EHEC-TWC)}
\label{sect_model}

We will first focus on the two-way channel and solve the problem at hand. We will then extend our solutions to the two-hop and multiple access models in Sections~\ref{sect_twohop} and \ref{sect_mac}, respectively.

\begin{figure}
   \includegraphics[width=0.9\linewidth]{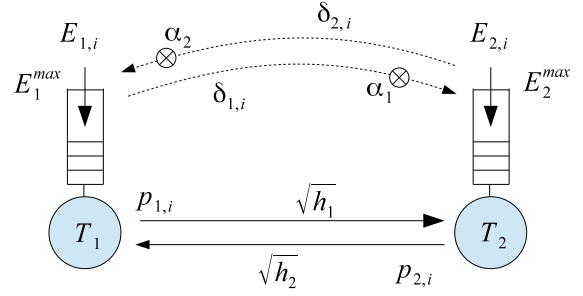}
   \centering
   \caption{Two-way channel with energy harvesting transmitters and energy cooperation.}
   \label{fig_twoway_model}
   \end{figure}

Consider the Gaussian two-way channel (TWC) \cite{shannon1961two} with two energy harvesting and energy cooperating (EHEC) nodes, $T_1$ and $T_2$, as shown in Fig.~\ref{fig_twoway_model}. Denoting the channel inputs by $X_k$ and the channel power gains by $h_k$, $k=1,2$, the channel outputs at nodes $T_1$ and $T_2$ after self interference cancellation are given by
\begin{align}
&Y_1=\sqrt{h_2}X_2+N_1, \\
&Y_2=\sqrt{h_1}X_1+N_2,
\end{align}
where $N_k$ is Gaussian noise with power $\sigma_k^2$ at node $T_k$, $k=1,2$. Each node cancels out its own contribution to the channel output, i.e., $T_k$ subtracts $X_k$ from $Y_k$, thus reducing the model to two parallel additive white Gaussian noise (AWGN) channels with channel power gains $h_1$ and $h_2$. We consider a static, i.e., time-invariant channel where $\sigma_k^2$ and $h_k$, $k=1,2$, remain constant throughout the transmission duration.

The communication session is divided into $1$sec long time slots\footnote{This choice is for simplicity. The results readily extend to slots with arbitrary length.}, indexed by $i=1,\dots,N$. Throughout the paper, we denote node indices by the first subscripts $k$, $j$ and $\ell$, and time slot indices by the second subscripts $i$ and $n$. In time slot $i$, node $T_k$, $k=1,2$, harvests $E_{k,i}$ units of energy, which it stores in its battery of size $E_k^{max}$. 
Within this time slot, $T_k$ transmits with average power $p_{k,i}$, which requires $p_{k,i}$ units of energy due to the unit slot length. In addition to harvesting energy, the nodes are also capable of transferring energy to each other. In time slot $i$, $T_k$ transfers $\delta_{k,i}$ units of energy to $T_j$, $j \neq k$. This transfer has an end-to-end efficiency of $\alpha_{k} \leq 1$, and $T_j$ receives $\alpha_{k}\delta_{k,i}$ units of energy as a result. 
The end-to-end transfer efficiency includes propagation loss, as well as other factors that scale linearly with the amount of energy transferred, e.g., circuit energy consumption at both parties.
The {\it power policy} of the network is defined as the collection of transmit powers and transferred energy values $\{p_{k,i},\delta_{k,i}\}$ for all $k$, $j$ and $i$.

The power policy of the network is constrained by the energy available to each node in each time slot. In particular, the energy remaining in the battery of node $T_k$ depends on the energy harvested at node $T_k$, the transmit power of the node, and the energy transferred to and received from node $T_j$ for $j \neq k$. We refer to the energy stored in the battery of $T_k$ at the end of time slot $i$ as the battery state $S_{k,i}$, which evolves as\footnote{We note that energy harvests and transfers are not instantaneous, but take place throughout the respective time slot. Since energy consumption via transmission is also distributed over the time slot, it is possible, and mathematically convenient, to express energy values as arriving to or departing from the node at the beginning of the slot.}
\begin{align}
\label{eqn_model_Bk}
S_{k,i}=\min \left\{ E_k^{max}, S_{k,i-1}+E_{k,i}-p_{k,i}-\delta_{k,i}+\alpha_{j}\delta_{j,i} \right\},
\end{align}
for $k=1,2$, with $S_{k,0}=0$. The initial charge of the batteries are introduced to the model through the energy harvests $E_{1,1}$ and $E_{2,1}$ in the first time slot. To ensure that the energy used by $T_k$ does not exceed the energy available at the node, we impose the set of {\em energy causality constraints} \cite{yang2012optimal, tutuncuoglu2012optimum}
\begin{align}
 	p_{k,i}+\delta_{k,i} \leq S_{k,i-1}+E_{k,i}+\alpha_{j}\delta_{j,i},
\end{align}
for $k=1,2$, $i=1,\dots,N$, which can equivalently be expressed as $S_{k,i} \geq 0$ for $k=1,2$ and $i=1,\dots,N$. These causality constraints imply that energy cannot be consumed, neither for transmission nor for transfer, before it is harvested or received. 

For average transmit powers $p_{1,i}$ and $p_{2,i}$ in time slot $i$, the sum-capacity of the Gaussian two-way channel in Fig.~\ref{fig_twoway_model} with full-duplex nodes is given by \cite{shannon1961two}
\begin{align}
\label{eqn_twoway_cs}
	C_S^{TWC} \left(p_{1,i},p_{2,i} \right) \! = \! \frac{1}{2} \log \left( 1+\tfrac{h_1 p_{1,i}}{\sigma_2^2} \right) \! + \! \frac{1}{2} \log \left( 1+\tfrac{h_2 p_{2,i}}{\sigma_1^2} \right).
\end{align}

We consider the offline problem (see also \cite{gurakan2012energy, gurakan2012two, gurakan2013energycoop, yang2012optimal, tutuncuoglu2012optimum, yang2012mac, gunduz2011two, tutuncuoglu2012sum}), where the energy harvests $E_{k,i}$ throughout the session are known at the beginning of the communication session. In addition to being applicable in networks with predictable energy arrivals, this approach also allows us to benchmark the performance limits of energy harvesting networks with energy cooperation. 

We consider the EHEC sum-throughput maximization problem for the TWC over transmit powers $p_{k,i}$ and energy transfers $\delta_{k,i}$, throughout a communication session of $N$ time slots, i.e.,
\begin{subequations}
\label{eqn_problem_twc}
\begin{align}
\label{eqn_problem_twc_a}
\underset{\{p_{k,i},\delta_{k,i}\}}{\max} ~ & \sum_{i=1}^N C_S^{TWC}(p_{1,i},p_{2,i}) \\
\label{eqn_problem_twc_b}
\mbox{s.t.} \quad ~ & S_{k,i} \geq 0, && \!\!\!\!\!\! k=1,2, ~ i=1,\dots,N, \\
\label{eqn_problem_twc_c}
 & p_{k,i} \geq 0, ~ \delta_{k,i} \geq 0, && \!\!\!\!\!\! k=1,2, ~ i=1,\dots,N,
\end{align}
\end{subequations}
where (\ref{eqn_problem_twc_b}) are the energy causality constraints, and (\ref{eqn_problem_twc_c}) are the non-negativity constraints for transmit power and transferred energy. We remark that similar problems that consider other criteria such as fairness can be formulated by updating the objective of (\ref{eqn_problem_twc}) accordingly.

We note that (\ref{eqn_problem_twc}) is either solved by all nodes in the network separately, or solved by one of the nodes and the output communicated to the others. This requires energy harvests and channel parameters to be shared between all energy harvesting transmitters in the system. The relatively small communication overhead, which is in the order of a few bytes per time slot, will be omitted in this work for the sake of simplicity.

Lastly, we remark that while linear end-to-end energy transfer losses are represented by $\{\alpha_k\}$, we do not explicitly account for other circuit or processing energy costs for energy cooperation. These can be incorporated into the optimization problem by embedding their cost models into (\ref{eqn_model_Bk}). Currently, experimental such models are being developed for specific transfer technologies, see for example \cite{kurs2007wireless} for strongly coupled magnetic resonance as the energy transfer technology.

\section{Properties of Optimal Policies for EHEC-TWC with Infinite Batteries}
\label{sect_properties}

We begin with the infinite battery case, $E_k^{max}=\infty$. In addition to providing insights about the optimal policy, this case is also a good approximation for systems where battery capacity is sufficiently large, harvested energy is sufficiently low, or transmission session is sufficiently brief with respect to harvesting period. The properties found in this section are extended to the finite battery case in Section~\ref{sect_finite_bat}. For this case, the battery state in (\ref{eqn_model_Bk}) can be rewritten as
\begin{align}
\label{eqn_properties_Bk}
S_{k,i} = \sum_{n=1}^{i} \left ( E_{k,n} - p_{k,n} + \alpha_{j} \delta_{j,n} - \delta_{k,n} \right),
\end{align}
for $k,j=1,2$ and $j \neq k$.

The problem in (\ref{eqn_problem_twc}) involves the joint optimization of transferred energy and transmit powers of the two nodes over $N$ time slots, i.e., $4N$ variables in total. In this section, we identify properties of optimal policies, the use of which helps us eliminate the additional complexity introduced by energy cooperation.

\subsection{Procrastinating Policies}
\label{sub_equivalent_procrastinating}

We first show that a subset of power policies, named {\it procrastinating policies}, includes at least one optimal policy.

\begin{definition}
\label{def_lazy}
	A power policy $\{p_{k,i},\delta_{k,i}\}$ is a procrastinating policy if it satisfies
	\begin{align}
		\label{eqn_lazy1}
		p_{k,i} - \alpha_{j} \delta_{j,i} \geq 0, \quad j,k=1,2, ~j \neq k, ~~i=1,\dots,N.
	\end{align}
\end{definition}

In each time slot, a procrastinating policy transfers energy from one node to the other only if all of the transferred energy is to be consumed for transmission immediately. This can be interpreted as the energy transferring nodes delaying energy transfers until the time slot they are immediately needed at the receiving end, hence the name {\em procrastinating} policies. In a procrastinating policy, energy transfers that are necessary for the feasibility of $p_{k,i}$ are postponed until the conditions in (\ref{eqn_lazy1}) are satisfied. The following lemma establishes the optimality of procrastinating policies.

\begin{lemma}
\label{lem_procrastinating}
	There exists at least one procrastinating policy that is a solution of (\ref{eqn_problem_twc}).
\end{lemma}
\begin{IEEEproof}
Let $\{p_{k,i}^*,\delta_{k,i}^*\}$ be an optimal policy which is not procrastinating, i.e., there exists $p_{k,n}^* < \alpha_{j} \delta_{j,n}^*$ for some $n$, $j$ and $k$, $j \neq k$. Starting from $i=1$, if $p_{k,i}^* < \alpha_{j} \delta_{j,i}^*$, set $\eta=p_{k,i}^* / (\alpha_{j} \delta_{j,i}^*)$, and update $\delta_{j,i+1}^*=\delta_{j,i+1}^*+(1-\eta) \delta_{j,i}^*$ and $\delta^*_{j,i}=\eta \delta_{j,i}^*$. This postpones excess transferred energy to the next time slot whenever (\ref{eqn_lazy1}) is violated. Note that the update in time slot $i$ only affects $S_{k,i}$ and $S_{j,i}$ in (\ref{eqn_problem_twc_b}), decreasing the former by $\alpha_{j} \delta_{j,i}^*-p_{k,i}^*$ and increasing the latter by $\delta_{j,i}^*-p_{k,i}^*/\alpha_{j}$. However, since $S_{k,i} \geq 0$ in the original policy and $\delta_{k,i}^*=0$ from Lemma~\ref{lem_unidirectional}, this change does not violate (\ref{eqn_problem_twc_b}). Repeating the updates for $i=2,\dots,N$ and $k=1,2$ yields a feasible procrastinating policy. Meanwhile, since $p_{k,i}^*$ is unchanged, the objective in (\ref{eqn_problem_twc_a}) is unchanged, and therefore the resulting procrastinating policy is also optimal.
\end{IEEEproof}

Lemma~\ref{lem_procrastinating} shows that by delaying energy transfers unless immediately required for transmission, any feasible transmit power policy $\{p_{k,i}\}$ can be realized with a procrastinating policy. Next, we utilize this property to decompose (\ref{eqn_problem_twc}) into two subproblems regarding the energy harvesting and energy cooperation aspects of the original problem.

\subsection{Decomposition to Energy Transfer and Power Allocation Problems}
\label{sub_equivalent_decomposition}

We define the {\it consumed power} $\bar{p}_{k,i}$ as the power drawn from the battery of node $T_k$, taking both transmission and transfers in time slot $i$ into consideration. This term is expressed as
\begin{align}
\label{eqn_equivalent_consumed}
\bar{p}_{k,i}&=p_{k,i} + \delta_{k,i} - \alpha_{j} \delta_{j,i}.
\end{align}
Note that by definition, consumed power can be negative. However, a procrastinating policy, as defined in Definition~\ref{def_lazy}, satisfies $\bar{p}_{k,i} \geq 0$ for all $i=1,\dots,N$ and $k=1,2$. We first present the following lemma.
\begin{lemma}
\label{lem_unidirectional}
	There exists an optimal policy which satisfies $\delta_{k,i}\delta_{j,i}=0$ for all $k,j=1,2$, $j \neq k$, $i=1,\dots,N$, i.e., energy transfer is never in both directions in a given time slot.
\end{lemma}
\begin{IEEEproof}
	Let $\{p_{k,i}^*,\delta_{k,i}^*\}$ be an optimal policy. Define $\tilde\delta_{k,i}=\max\{\delta_{k,i}^*-\delta_{j,i}^*,0\}$ for all $k,j=1,2$, $j\neq k$, and $n=1,\dots,N$, which satisfies $\tilde\delta_{k,i}\tilde\delta_{j,i}=0$. With these energy transfers, $S_{k,i}$ in (\ref{eqn_properties_Bk}) increases for all $k$ and $i$, and therefore the procrastinating policy $\{p_{k,i}^*,\tilde\delta_{j,i}\}$ is feasible. Since $p_{k,i}^*$ are unchanged, it also yields the same objective as $\{p_{k,i}^*,\delta_{j,i}^*\}$ and is therefore optimal.
\end{IEEEproof}

Lemma~\ref{lem_unidirectional} is a natural consequence of transfer efficiencies being less than 100\%, and is intuitively pleasing. As a consequence of the lemma, we can restrict our attention to policies satisfying the lemma without loss of optimality. Hence, for procrastinating policies satisfying the lemma, the non-negativity constraints $p_{k,i}\geq 0$ in (\ref{eqn_problem_twc_c}) are equivalent to $\bar{p}_{k,i} \geq \delta_{k,i}$, $i=1,\dots,N$, $k=1,2$. Restricting the feasible set of (\ref{eqn_problem_twc}) to procrastinating policies satisfying Lemma~\ref{lem_unidirectional}, without loss of optimality, we rewrite (\ref{eqn_problem_twc}) as
\begin{subequations}
\label{eqn_problem2}
\begin{align}
\label{eqn_problem2_a}
	\underset{\{\bar{p}_{k,i}, \delta_{k,i}\}}{\max} ~
	& \sum_{i=1}^{N}  C_S^{TWC} \bigg( \bigg[~ \bar{p}_{k,i}+ \alpha_{k}\delta_{k,i}-\delta_{j,i} ~\bigg] \bigg) \\
\label{eqn_problem2_b}
	\mbox{s.t.} \quad ~ & \sum_{n=1}^{i}E_{k,n} - \bar{p}_{k,n} \geq 0, ~~ k=1,2, ~ i=1,\dots,N, \\
\label{eqn_problem2_c}
	& \bar{p}_{k,i} \geq \delta_{k,i}, ~ \delta_{k,i} \geq 0, ~~ k=1,2, ~ i=1,\dots,N.
\end{align}
\end{subequations} 
Here, $[p_{k,i}]=(p_{1,i},p_{2,i})$ denotes both parameters of $C_S^{TWC}$, which are found by substituting $k=1,2$ and $j \neq k$. Note that the constraints in (\ref{eqn_problem_twc_b}), which include both energy transfers $\delta_{k,i}$ and transmit powers $p_{k,i}$, are replaced with (\ref{eqn_problem2_b})-(\ref{eqn_problem2_c}) where energy transfers and consumed powers are now decoupled. Furthermore, the $i$th summation term in the objective (\ref{eqn_problem2_a}) depends only on the variables for the respective time slot $i$. Hence, (\ref{eqn_problem2}) can be decomposed as
\begin{subequations}
\label{eqn_problem_PA}
\begin{align}
\label{eqn_problem_pa_a}
	\underset{\{\bar{p}_{k,i}\}}{\max} ~& \sum_{i=1}^{N} R(\bar{p}_{1,i},\bar{p}_{2,i}) \\
\label{eqn_problem_pa_b}
	\mbox{s.t.} ~& \sum_{n=1}^{i}E_{k,n} - \bar{p}_{k,n} \geq 0, ~~ k=1,2,~ i=1,\dots,N, \\
\label{eqn_problem_pa_c}
	& \hspace{0.72in} \bar{p}_{k,i} \geq 0, ~~ k=1,2,~ i=1,\dots,N,
\end{align}
\end{subequations} 
where $R(\bar{p}_{1,i},\bar{p}_{2,i})$ is the {\em per-slot sum-rate} for consumed powers $\bar{p}_{1,i}$ and $\bar{p}_{2,i}$, defined as
\begin{subequations}
\label{eqn_problem_ET}
\begin{align}
\label{eqn_problem_et_a}
R(\bar{p}_{1,i},\bar{p}_{2,i})= & \max_{\delta_{1,i},\delta_{2,i}}
C_S^{TWC} \left( \big[~ \bar{p}_{k,i}+ \alpha_k\delta_{k,i}-\delta_{j,i} ~\big] \right) \\
\label{eqn_problem_et_b}
& ~~ \mbox{s.t.} ~~ \bar{p}_{k,i} \geq \delta_{k,i}, ~~ \delta_{k,i} \geq 0, ~k=1,2.
\end{align}
\end{subequations}
Note that (\ref{eqn_problem_ET}) yields the optimal energy transfers $\delta_{k,i}$ within a single time slot $i$ for a fixed pair of consumed powers $(\bar{p}_{1,i},\bar{p}_{2,i})$. Being separated from $\delta_{k,i}$, (\ref{eqn_problem_PA}) finds the optimal allocation of consumed powers $\bar{p}_{k,i}$, $i=1,\dots,N$ throughout the transmission. This decomposition implies that the power transfer optimization can be performed separately and in a slot-by-slot basis, i.e., the optimal energy transfers $\delta_{k,i}$ can be found using only the consumed powers $\bar{p}_{k,i}$ in the same time slot.

\begin{lemma}
\label{lem_concavity_of_r}
	$R(\pi_1,\pi_2)$, is jointly concave in $\pi_{1}$ and $\pi_{2}$.
\end{lemma}
\begin{IEEEproof}
The proof can be found in Appendix~\ref{app_proof_concavity}.
\end{IEEEproof}

As a result of Lemma~\ref{lem_concavity_of_r}, the consumed power allocation problem in (\ref{eqn_problem_PA}) is a convex program. Furthermore, the constraints in (\ref{eqn_problem_pa_b}) and (\ref{eqn_problem_pa_c}) are separable among transmitters $k=1,2$, and hence a block coordinate descent (alternating maximization) algorithm that alternates between $\{\bar{p}_{1,i}\}$ and $\{\bar{p}_{2,i}\}$ converges to the optimal policy \cite{bertsekas1999nonlinear}. In particular, at each iteration, a single transmitter problem with a concave objective function and linear energy causality constraints is solved. The iterations evolve, alternating over the optimized variables, until the policies converge. Namely, we solve
\begin{subequations}
\label{eqn_problem_PA_iter}
\begin{align}
\label{eqn_problem_pa_iter_a}
\underset{\{\bar{p}_{k,i}\}}{\max} & ~~ \sum_{i=1}^{N} R(\bar{p}_{1,i},\bar{p}_{2,i}) ~~~~~ &\\
\label{eqn_problem_pa_iter_b}
\mbox{s.t.}  & ~~ \sum_{n=1}^{i}E_{k,n} - \bar{p}_{k,n} \geq 0, & i=1,\dots,N,\\
\label{eqn_problem_pa_iter_c}
& ~~ \hspace{0.72in} \bar{p}_{k,i} \geq 0, & i=1,\dots,N,
\end{align}
\end{subequations}  
for a fixed $k$ at each iteration, alternating between $k=1$ and $k=2$, while $\{\bar{p}_{j,i}\}$, $j \neq k$, is held constant. Note that (\ref{eqn_problem_PA_iter}) differs from its counterpart without energy cooperation \cite{yang2012optimal} only in the rate function $R(\bar{p}_{1,i},\bar{p}_{2,i})$. However, $\bar{p}_{j,i}$, $j \neq k$, may change in time, and hence the solution to each iteration step in (\ref{eqn_problem_PA_iter}) is not the constant power policy in \cite{yang2012optimal}. Instead, it can be found using a generalized directional water-filling algorithm, as we will describe next.

\section{Optimal Policy for the EHEC-TWC with Infinite Batteries}
\label{sect_twc}

The decomposition in (\ref{eqn_problem_PA})-(\ref{eqn_problem_ET}) simplifies the analysis of the problem by separating the power allocation problem from energy transfer variables $\{\delta_{k,i}\}$, and calculating optimal energy transfers in a slot-by-slot basis. We first solve the energy transfer problem within a single slot, i.e., (\ref{eqn_problem_ET}), which we then substitute in (\ref{eqn_problem_PA_iter}) to solve (\ref{eqn_problem_PA}).

\subsection{Optimal Energy Transfers for the EHEC-TWC}
\label{sub_twc_et}

Consider time slot $i$ first. We focus on the two subsets of the feasible space of (\ref{eqn_problem_ET}), namely those satisfying $\delta_{1,i}=0$ and $\delta_{2,i}=0$, one of which contains an optimal policy as implied by Lemma~\ref{lem_unidirectional}. We solve (\ref{eqn_problem_ET}) for these subsets, and choose the maximum of the two. For the policies satisfying $\delta_{j,i}=0$, the solution to (\ref{eqn_problem_ET}) is found as
\begin{align}
\label{eqn_twc_et_solution}
	\delta^*_{k,i}=\min \left\{\bar{p}_{k,i}, \frac{1}{2} \left[  \left( \frac{\sigma_j^2}{h_k}+\bar{p}_{k,i} \right)-\frac{1}{\alpha_k}\left( \frac{\sigma_k^2}{h_j}+\bar{p}_{j,i} \right) \right]^+ \right\}
\end{align}
for $k \neq j$, where $[x]^+$ denotes $\max\{0,x\}$. This yields two optimal transfer candidates, $\delta^*_{1,i}$ and $\delta^*_{2,i}$, each requiring the other to be zero. Note that the case where both candidates are positive, i.e., $\delta^*_{1,i}>0$ and $\delta^*_{2,i}>0$, requires $\alpha_{1}\alpha_{2}>1$, which is not possible since $\alpha_1,\alpha_2 \leq 1$ by definition. Hence, at least one of the two candidates is always zero, and (\ref{eqn_twc_et_solution}) immediately gives the solution to (\ref{eqn_problem_ET}). The per-slot sum-rate achieved by the optimal energy transfer policy, corresponding to $R(\bar{p}_{1,i},\bar{p}_{2,i})$ in (\ref{eqn_problem_PA}) and (\ref{eqn_problem_ET}), is then expressed as
\begin{align}
\nonumber
	& R(\bar{p}_{1,i},\bar{p}_{2,i}) = \\
\label{eqn_twoway_R}
	& \begin{cases}
	C_S^{TWC}(\bar{p}_{1,i},\bar{p}_{2,i}), \hfill \delta^*_{1,i}=\delta^*_{2,i}=0, \\
	\log \left( \frac{\sqrt{\alpha_1 h_1 h_2}}{2\sigma_1\sigma_2}
		\left( \left( \frac{\sigma_2^2}{h_1}+\bar{p}_{1,i} \right)+\frac{1}{\alpha_1}\left( 
		\frac{\sigma_1^2}{h_2}+\bar{p}_{2,i} \right) \right) \right), \qquad \quad \\
		\null\hfill 0<\delta^*_{1,i}<\bar{p}_{1,i}, \\
	\log \left( \frac{\sqrt{\alpha_2 h_1 h_2}}{2\sigma_1\sigma_2}
		\left( \left( \frac{\sigma_1^2}{h_2}+\bar{p}_{2,i} \right)+\frac{1}{\alpha_2}\left( 
		\frac{\sigma_2^2}{h_1}+\bar{p}_{1,i} \right) \right) \right), \\
		\null\hfill 0<\delta^*_{2,i}<\bar{p}_{2,i}, \\
	\frac{1}{2} \log \left(1+\frac{h_2}{\sigma_1^2}\left(\bar{p}_{2,i}+\alpha_{1}\bar{p}_{1,i}\right)\right),
		\hfill 0<\delta^*_{1,i}=\bar{p}_{1,i}, \\
	\frac{1}{2} \log \left(1+\frac{h_1}{\sigma_2^2}\left(\bar{p}_{1,i}+\alpha_{2}\bar{p}_{2,i}\right)\right),
		\hfill 0<\delta^*_{2,i}=\bar{p}_{2,i}.	
	\end{cases}	
\end{align}

\subsection{Optimal Power Allocation for the EHEC-TWC}
\label{sub_twc_pa}

Substituting (\ref{eqn_twoway_R}) in (\ref{eqn_problem_PA_iter}), it remains to solve for the optimal $\{\bar{p}_{k,i}\}$ by iterating between $\bar{p}_{1,i}$ and $\bar{p}_{2,i}$. We now show that the solution to each iteration admits a {\em generalized directional water-filling} interpretation, and consequently (\ref{eqn_problem_twc}) can be solved using the generalized iterative directional water-filling algorithm \cite{yang2012mac,tutuncuoglu2012sum}.

As shown in Lemma~\ref{lem_concavity_of_r}, (\ref{eqn_problem_PA_iter}) is a convex program with affine constraints. We also remark that $R(\pi_1,\pi_2)$ in (\ref{eqn_twoway_R}) is a continuously differentiable function in both $\pi_1$ and $\pi_2$. Hence, the KKT optimality conditions of (\ref{eqn_problem_PA_iter}) are necessary and sufficient for optimality, and are found as
\begin{align}
\label{eqn_inf_stat_1}
	-\frac{d R(\bar{p}_{1,i},\bar{p}_{2,i})}{d \bar{p}_{k,i}} + \sum_{n=i}^N \lambda_{k,n} - \tau_{k,i} &= 0, \\
\label{eqn_inf_cs_1}
	\lambda_{k,i} S_{k,i} =0, \quad \tau_{k,i} \bar{p}_{k,i} =0,
\end{align}
for $k,j=1,2$, $j \neq k$, and $i=1,\dots,N$. Here, $\lambda_{k,i} \geq 0$ and $\tau_{k,i} \geq 0$ are the Lagrange multipliers for the constraints in (\ref{eqn_problem_pa_iter_b}) and (\ref{eqn_problem_pa_iter_c}), respectively. For the optimal $\{\bar{p}_{k,i}\}$ solving (\ref{eqn_problem_PA_iter}), there exists a set of non-negative Lagrangian multipliers that satisfy the conditions in (\ref{eqn_inf_stat_1})-(\ref{eqn_inf_cs_1}), and vice versa. We define
\begin{align}
\label{eqn_inf_waterlevel}
	v_{k,i}= \left( \frac{d R(\bar{p}_{1,i},\bar{p}_{2,i})}{d \bar{p}_{k,i}} \right)^{-1}.
\end{align}
From (\ref{eqn_inf_cs_1}), we observe that whenever $\bar{p}_{k,i}>0$, we have $\tau_{k,i}=0$. In this case, from (\ref{eqn_inf_stat_1}), we see that optimal $\{v_{k,i}\}$ are constant in $i$ unless $\lambda_{k,i}>0$. Meanwhile, a positive $\lambda_{k,i}$ is only possible when $S_{k,i}=0$, i.e., the battery is empty. Moreover, since $\lambda_{k,i}$ are non-negative, optimal $\{v_{k,i}\}$ are non-decreasing in $i$. Due to this behavior, we refer to $\{v_{k,i}\}$ as {\em generalized water levels}, and utilize the directional water-filling interpretation in \cite{ozel2011fading} to find the optimal water levels.

In particular, in each iteration for $T_k$, transmit powers and water levels are initialized by setting $\bar{p}_{k,i}=E_{k,i}$ for all $i=1,\dots,N$. If the water levels satisfy $v_{k,i}>v_{k,i+1}$ for some $i$, this results in water (energy) flow from slot $i$ to slot $i+1$, which is achieved by decreasing $\bar{p}_{k,i}$ and increasing $\bar{p}_{k,i+1}$ by the same amount. The flow stops when the two water levels are equalized, or when $\bar{p}_{k,i}=0$. In the latter case, (\ref{eqn_inf_stat_1}) is satisfied via some $\tau_{k,i}>0$, which is feasible since $\bar{p}_{k,i}=0$. Flow in the reverse direction, e.g., from time slot $i+1$ to $i$, is not allowed since it violates the energy causality constraints in (\ref{eqn_problem_twc_b}). The algorithm terminates when water levels do not permit any water flow.

The generalized directional water-filling algorithm above is then repeated at each iteration. For the iterations on $\bar{p}_{k,i}$, where $\bar{p}_{j,i}$, $j \neq k$ is kept constant, the water levels are given by
\begin{align}
\label{eqn_wlevel_1}
	v_{k,i}= 	
	\begin{cases}
	2\left( \frac{\sigma_j^2}{h_k}+\bar{p}_{k,i} \right), 
		& \delta_{k,i}^*=\delta_{j,i}^*=0, \\
	\left( \frac{\sigma_j^2}{h_k}+\bar{p}_{k,i}\right)
		+\tfrac{1}{\alpha_k}\left(\frac{\sigma_k^2}{h_j}+\bar{p}_{j,i}\right),
		& 0<\delta^*_{k}<\bar{p}_k, \\
	\left( \frac{\sigma_j^2}{h_k}+\bar{p}_{k,i}\right)
		+\alpha_j \left(\frac{\sigma_k^2}{h_j}+\bar{p}_{j,i}\right),
		& 0<\delta^*_{j}<\bar{p}_j, \\
	2 \left( \bar{p}_{k,i}+\frac{1}{\alpha_k}\left(\frac{\sigma_k^2}{h_j}+\bar{p}_{j,i} \right) \right),
		& 0<\delta^*_k=\bar{p}_k, \\
	2 \left(\frac{\sigma_j^2}{h_k}+\bar{p}_{k,i}+{\alpha_j}\bar{p}_{j,i} \right),
		& 0<\delta^*_j=\bar{p}_j. \\
	\end{cases}	
\end{align}

\begin{figure}
   \includegraphics[width=\linewidth]{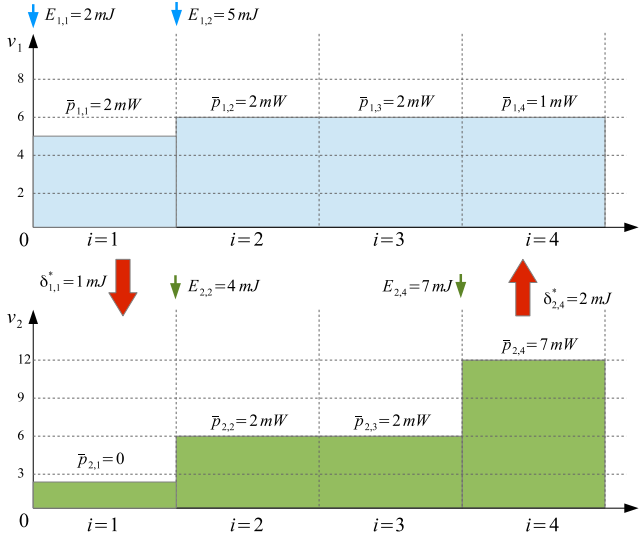}
   \centering
   \caption{Optimal water levels found by iterative generalized water-filling with water levels in (\ref{eqn_wlevel_1}).}
   \label{fig_wf_twc}
   \end{figure}

We present an example of the directional water-filling algorithm for $N=4$~time slots of length $1$~sec and $\alpha_{1}=\alpha_{2}=0.5$ in Fig.~\ref{fig_wf_twc}. Energy arrivals to the two nodes are $E_1=[2,5,0,0]$~mJ and $E_2=[0,4,0,7]$~mJ. The final (equilibrium) water levels are shown in blue for node $T_1$, and in green for node $T_2$, for $h_1=h_2=-100$~dB, and $\sigma_1^2=\sigma_2^2=10^{-13}$~W for a 1~MHz bandwidth. Observe that in the first time slot, $\bar{p}_{1,1}=2$~mW and $\bar{p}_{2,1}=0$ yields the optimal energy transfers $\delta^*_{1,1}=1$~mJ and $\delta^*_{2,1}=0$, i.e., node $T_1$ transfers $1$~mJ of energy to node $T_2$ as indicated with the red arrow. The energy transfer candidates for time slots $i=2,3$ are zero, and no energy is transferred. In the last time slot, the optimal energy transfer rate is found as $\delta^*_{1,4}=0$ and $\delta^*_{2,4}=2$~mJ, i.e., the energy transfer is from node $T_2$ to node $T_1$. With the final water levels in the figure, no further water flow is feasible for either node.

\section{The EHEC Two-Hop Channel}
\label{sect_twohop}

\begin{figure}
   \includegraphics[width=0.9\linewidth]{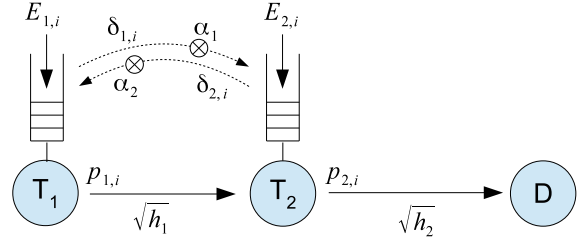}
   \centering
   \caption{The two-hop channel model with energy harvesting transmitters and relay, and energy cooperation.}
   \label{fig_twohop_model}
   \end{figure}

We next consider a two-hop channel (THC) with infinite-sized batteries as a simple example of a multi-hop setting as done in \cite{gurakan2012energy}, extended to bi-directional energy transfers. The channel model is as shown in Fig.~\ref{fig_twohop_model}. For this case, we denote the transmit power of the source node by $p_{1,i}$, the relay node by $p_{2,i}$, and the source-relay and relay-destination channel power gains by $h_1$ and $h_2$, respectively. The messages are delay constrained, and the relay node $T_2$ needs to forward all received messages immediately to the destination.
As such, the relay does not have a data buffer, and departs packets in the same time slot they are received.
The source and the relay are both capable of energy transfer. The sum-capacity for this channel with a full-duplex relay is given by
\begin{align}
\nonumber
& C_S^{THC}(p_{1,i},p_{2,i})= \min \Bigg\{ \frac{1}{2} \log \left( 1+ \frac{h_1 p_{1,i}}{\sigma_2^2} \right),\\
\label{eqn_twohop_rate} 
& \hspace{1.5in} \frac{1}{2} \log \left( 1+ \frac{h_2 p_{2,i}}{\sigma_1^2} \right) \Bigg\}.
\end{align}
Note that as in the two-way model, $C_S^{THC}$ is jointly concave in $p_{1,i}$ and $p_{2,i}$ since it is the minimum of two jointly concave functions. Hence, the throughput maximization problem for this channel also satisfies Lemma~\ref{lem_procrastinating}, and therefore allows the decomposition in (\ref{eqn_problem_PA})-(\ref{eqn_problem_ET}).

\subsection{Optimal Energy Transfers for the EHEC-THC}
\label{sub_twohop_et}

Given the capacity expression in (\ref{eqn_twohop_rate}), the two-hop version of (\ref{eqn_problem_ET}) can be written as
\begin{subequations}
\label{eqn_twohop_ET}
\begin{align}
\label{eqn_twohop_et_a}
\underset{\delta_{1,i},\delta_{2,i}}{\max} & \min \left\{ 
	\tfrac{h_1}{\sigma_2^2} \left( \bar{p}_{1,i} \!+\! \alpha_{1}\delta_{1,i} \!-\! \delta_{2,i} \right), 
	\tfrac{h_2}{\sigma_1^2} \left( \bar{p}_{2,i} \!+\! \alpha_{2}\delta_{2,i} \!-\! \delta_{1,i} \right) \right\} \\
\label{eqn_twohop_et_b}
\mbox{s.t.} ~ & ~  \bar{p}_{k,i} \geq \delta_{k,i}, ~~ \delta_{k,i} \geq 0, ~~k=1,2.
\end{align}
\end{subequations}
The objective is the minimum of two linear functions, and the two terms of the minimum change in opposite directions with $\delta_{1,i}$ or $\delta_{2,i}$. Hence, the optimal is attained when the two terms are equal, if feasible. Solving (\ref{eqn_twohop_ET}) for $\delta_{1,i}$ and $\delta_{2,i}$ satisfying
\begin{equation}
\label{eqn_twohop_et_solution}
\frac{h_1}{\sigma_2^2} \left( \bar{p}_{1,i}+ \alpha_{1}\delta_{1,i}-\delta_{2,i} \right)
= \frac{h_2}{\sigma_1^2} \left( \bar{p}_{2,i}+ \alpha_{2}\delta_{2,i}-\delta_{1,i} \right)
\end{equation}
yields the energy transfers
\begin{subequations}
\label{eqn_marc_delta_opt_single}
\begin{align}
\delta_{1,i} &= \left[ \frac{\sigma_1^2 h_1\bar{p}_{1,i} - \sigma_2^2 h_2 \bar{p}_{2,i}}{\alpha_{1}\sigma_2^2 h_2 + \sigma_1^2 h_1} \right] ^+ , \\
\delta_{2,i} &= \left[ \frac{\sigma_2^2 h_2\bar{p}_{2,i} - \sigma_1^2 h_1 \bar{p}_{1,i}}{\alpha_{2}\sigma_1^2 h_1 + \sigma_2^2 h_2} \right] ^+ ,
\end{align}
\end{subequations}
where $[x]^+$ denotes $\max\{0,x\}$. Since $\alpha_1,\alpha_2 \geq 0$, these energy transfer values are feasible and therefore optimal. Observe that due to (\ref{eqn_marc_delta_opt_single}), the difference of received powers, i.e., $\frac{h_1}{\sigma_2^2}\bar{p}_{1,i} - \frac{h_2}{\sigma_1^2} \bar{p}_{2,i}$ determines the direction of energy transfer, and the transferred energy is nonzero unless the two received powers are equal.

\subsection{Optimal Power Allocation for the EHEC-THC}
\label{sub_twohop_pa}

Substituting the optimal values in (\ref{eqn_marc_delta_opt_single}) into the power allocation problem in (\ref{eqn_problem_PA}) yields
\begin{subequations}
\label{eqn_marc_subproblem}
\begin{align}
\label{eqn_marc_subproblem_a}
\underset{\{\bar{p}_{k,i}\}}{\max} & ~ \sum_{i=1}^N \log \left( 1+ h_1 h_2
	\min \left\{ \tfrac{\bar{p}_{1,i}+\alpha_{2} \bar{p}_{2,i}}{\alpha_{2}\sigma_1^2h_1+\sigma_2^2h_2}, 
	~ \tfrac{\alpha_{1} \bar{p}_{1,i}+\bar{p}_{2,i}}{\sigma_1^2h_1+\alpha_{1}\sigma_2^2h_2} \right\} \right) \\
\label{eqn_marc_subproblem_b}
\mbox{s.t.} & ~ \sum_{n=1}^{i} E_{k,n} \!-\! \bar{p}_{k,n} \geq 0, ~\bar{p}_{k,i} \!\geq\! 0, ~~ k \!= \!1,2,~i \!=\! 1,\dots,N.
\end{align}
\end{subequations} 
Due to the convexity of the problem, the generalized iterative water-filling algorithm in Section~\ref{sub_twc_pa} can also be used for (\ref{eqn_marc_subproblem}) by solving (\ref{eqn_marc_subproblem}) iteratively in $\{\bar{p}_{1,i}\}$ and $\{\bar{p}_{2,i}\}$. In this model, the generalized water levels are found for the $\{\bar{p}_{k,i}\}$ iteration, keeping $\{\bar{p}_{j,i}\}$, $j\neq k$ constant, as 
\begin{equation}
\label{eqn_thc_wlevel}
v_{k,i} \!=\!
\begin{cases} 
\bar{p}_{k,i} + \alpha_j \bar{p}_{j,i} + \left( \frac{\sigma_j^2}{h_k} + \frac{\alpha_j \sigma_k^2}{h_j} \right), 
	& \!\!\! \sigma_k^2 h_k \bar{p}_{k,i} < \sigma_j^2 h_j \bar{p}_{j,i}, \\
\bar{p}_{k,i} + \frac{\bar{p}_{j,i}}{\alpha_k} + \left(\frac{\sigma_j^2}{h_k} + \frac{\sigma_k^2}{\alpha_k h_j} \right), 
	& \!\!\! \sigma_k^2 h_k \bar{p}_{k,i} \geq \sigma_j^2 h_j \bar{p}_{j,i}.
\end{cases}
\end{equation}
We remark that the water levels are linear in transmit powers, and therefore the algorithm resembles conventional water-filling \cite{goldsmith1997capacity}. In the iteration on $\{\bar{p}_{k,i}\}$, the consumed powers $\{\bar{p}_{j,i}\}$, $j \neq k$ are kept constant, which introduces a base level over which water-filling is performed. The first terms in (\ref{eqn_thc_wlevel}) are consumed powers, the second terms are base levels due to the other transmitter, and the third terms are constant.

We further remark that water levels $v_{1,i}$ and $v_{2,i}$ are linearly related, with ratio $\alpha_{2}$ or $\alpha_{1}$ depending on the direction of energy transfer. As a consequence, unless $\alpha_1=\alpha_2=1$, if the water levels in two consecutive time slots are equal for both transmitters, the direction of energy transfer must remain the same as well. An insight that can be drawn from this observation is that the direction of energy transfer remains unchanged in time unless the water levels change for one of the nodes, which only occurs when the respective node is out of energy.

\begin{figure}
\centering
	\includegraphics[width=0.9\linewidth]{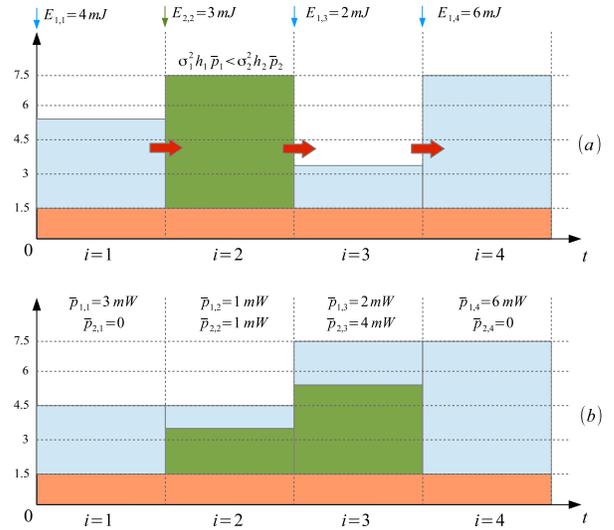}
	\caption{Directional water-filling (a) initial levels and (b) levels after water flow, for an example setting with $N=4$. The blue, green and orange areas correspond to source consumption $\bar{p}_1$, relay consumption $\bar{p}_2$ and the constant term in (\ref{eqn_thc_wlevel}), respectively.} 
\label{fig_waterfilling}
\end{figure}

Combining these two remarks, we observe that the generalized directional water-filling algorithm has an intuitive two-fluid interpretation for the two-hop channel. Namely, we can solve (\ref{eqn_marc_subproblem}) by considering $\bar{p}_{1,i}$ and $\bar{p}_{2,i}$ as levels of {\em two immiscible fluids}, and scaling these fluids appropriately in each iteration while performing directional water-filling based on the total water level. The analogy is even more apparent in the uni-directional energy transfer case, where $\alpha_{2}=0$. This gives $v_{2,i}=\infty$ whenever $\sigma_1^2 h_1 \bar{p}_{1,i} < \sigma_2^2 h_2 \bar{p}_{2,i}$, thus restricting the solution to $\sigma_1^2 h_1 \bar{p}_{1,i} \geq \sigma_2^2 h_2 \bar{p}_{2,i}$. With this restriction, we have $v_{2,i}=\alpha_{1}v_{1,i}$, and therefore no iteration is necessary. The optimal consumed powers are found as the resulting water levels when both fluids are allowed to flow while satisfying the condition $\sigma_1^2 h_1 \bar{p}_{1,i} \geq \sigma_2^2 h_2 \bar{p}_{2,i}$. An example to this two-fluid water-filling is depicted in Fig.~\ref{fig_waterfilling} for $\alpha_{1}=0.5$, $\alpha_{2}=0$, $E_{1,i}=[4,0,2,6]$~mJ, $E_{2,i}=[0,3,0,0]$~mJ, and the same channel parameters in Fig.~\ref{fig_wf_twc}. Note that in this example, water flow for node $T_2$ (green) from $i=2$ to $i=3$ occurs, even against the level gradient, until the condition $\sigma_1^2 h_1 \bar{p}_{1,2} \geq \sigma_2^2 h_2 \bar{p}_{2,2}$ is satisfied.

\section{The EHEC Multiple Access Channel}
\label{sect_mac}

\begin{figure}
   \includegraphics[width=0.9\linewidth]{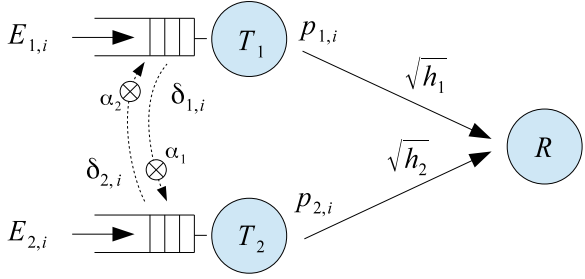}
   \centering
   \caption{K-transmitter multiple access channel with energy harvesting transmitters and energy cooperation.}
   \label{fig_mac_model}
   \end{figure}

In this section, we extend the results in Sections~\ref{sect_properties} and \ref{sect_twc} to the Gaussian multiple access channel (MAC) with infinite-sized batteries, shown in Fig.~\ref{fig_mac_model}. The sum-capacity for this channel in time slot $i$ is given by
\begin{align}
\label{eqn_mac_capacity}
	C_S^{MAC}(p_{1,i},p_{2,i}) = \frac{1}{2} \log \left( 1+ \frac{h_1 p_{1,i}}{\sigma_2^2} + \frac{h_2 p_{2,i}}{\sigma_1^2} \right).
\end{align}
The corresponding sum-throughput maximization problem, i.e., the MAC version of (\ref{eqn_problem_twc}), is
\begin{subequations}
\label{eqn_problem_mac}
\begin{align}
\label{eqn_problem_mac_a}
\underset{\{p_{k,i},\delta_{k,i}\}}{\max} & ~ \sum_{i=1}^N C_S^{MAC}(p_{1,i},p_{2,i})\\
\label{eqn_problem_mac_b}
\mbox{s.t.} \quad & ~ S_{k,i} \!\geq\! 0, ~ p_{k,i} \!\geq\! 0, ~ \delta_{k,i} \!\geq\! 0, ~~ k \!=\! 1,2, ~ i \!=\! 1,\dots,N.
\end{align}
\end{subequations}
Since  $C_S^{MAC}$ is also jointly concave in $p_{1,i}$ and $p_{2,i}$, the MAC sum-throughput maximization problem also satisfies Lemma~\ref{lem_procrastinating}, yielding the decomposition in (\ref{eqn_problem_PA})-(\ref{eqn_problem_ET}). In the following subsections, we address the energy transfer and power allocation subproblems for the MAC.

\subsection{Optimal Energy Transfers for the EHEC-MAC}
\label{sub_mac_optimal}

Substituting the consumed powers in (\ref{eqn_equivalent_consumed}) into (\ref{eqn_problem_mac}) yields
\begin{subequations}
\label{eqn_problem_mac_pa}
\begin{align}
\label{eqn_problem_mac_pa_a}
	\underset{\{\bar{p}_{k,i}\}}{\max} ~& 
	\sum_{i=1}^{N}  C_S^{MAC} \left( \bigg[~ \bar{p}_{k,i}+ \alpha_{k}\delta_{k,i}-\delta_{j,i} ~\bigg] \right) \\
\label{eqn_problem_mac_pa_b}
	\mbox{s.t.} ~ & \sum_{n=1}^{i}E_{k,n} - \bar{p}_{k,n} \geq 0, \quad k=1,2,~ i=1,\dots,N,
\end{align}
\end{subequations}
where $\delta_{1,i}$ and $\delta_{2,i}$ are found as the solution to the energy transfer problem 
\begin{subequations}
\label{eqn_problem_mac_et}
\begin{align}
\label{eqn_problem_mac_et_a}
\max &~ \delta_{1,i} \left( \alpha_{1}\frac{h_2}{\sigma_1^2}-\frac{h_1}{\sigma_2^2} \right) 
	+ \delta_{2,i} \left( \alpha_{2}\frac{h_1}{\sigma_2^2}-\frac{h_2}{\sigma_1^2} \right) \\
\label{eqn_problem_mac_et_b}
\mbox{s.t.} &~ 0 \leq \delta_{k,i} \leq \bar{p}_{k,i},~~k=1,2.
\end{align}
\end{subequations}
Note that (\ref{eqn_problem_mac_et}) is a linear program, with the optimal achieved at a corner of the rectangle defined by (\ref{eqn_problem_mac_et_b}). The optimal policy is to choose $\delta_{k,i}=\bar{p}_{k,i}$ if $\alpha_{k}\sigma_j^2 h_j>\sigma_k^2 h_k$, and choose $\delta_{k,i}=0$ otherwise. Consequently, the allocated power at $T_k$ is entirely transferred to $T_j$ if $\alpha_{k}\sigma_j^2 h_j>\sigma_k^2 h_k$, or is entirely used for transmission if $\alpha_{k}\sigma_j^2 h_j \leq \sigma_k^2 h_k$. This also implies that energy transfers only depend on the channel parameters, and hence the optimal energy transfer direction remains the same throughout the transmission. 
As a result, uni-directional energy transfer is sufficient from one user to the other in the direction determined by the channels and their transfer efficiency values.

\subsection{Optimal Power Allocation for the EHEC-MAC}
\label{sub_mac_pa}

\begin{figure}
   \includegraphics[width=0.9\linewidth]{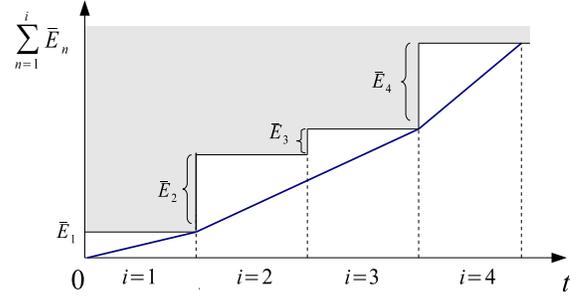}
   \centering
   \caption{Optimal sum-power policy for an energy cooperating MAC for a deadline of $N=4$ time slots.}
   \label{fig_stairs_mac}
   \end{figure}

The analysis in Section~\ref{sub_mac_optimal} reveals that in the optimal policy, either no energy transfer occurs, or one node transfers all of its energy to the other. In the former case, we get an energy harvesting MAC without energy transfers, the sum-capacity of which was found in \cite{yang2012mac}. The problem is solved by combining harvested energy in a single pool, thus reducing the problem to the single link power allocation problem in \cite{yang2012optimal}. In the latter case, let $\alpha_{2}\sigma_1^2 h_1>\sigma_2^2 h_2$ without loss of generality. Then, the optimal energy transfers in Section~\ref{sub_mac_optimal} yield the water-levels
\begin{align}
\label{eqn_mac_wlevel}
v_{1,i}=\frac{\sigma_2^2}{h_1}+\bar{p}_{1,i}+\alpha_2\bar{p}_{2,i}, \quad v_{2,i}=\frac{\sigma_2^2}{h_1 \alpha_2}+\frac{\bar{p}_{1,i}}{\alpha_2}+\bar{p}_{2,i}.
\end{align}
Note that $v_{1,i}=\alpha_2 v_{2,i}$. In this case, we can equivalently consider the policy of node $T_1$ only, and transfer all energy harvested by node $T_2$ immediately to node $T_1$. We establish this by scaling $\{E_{2,i}\}$ with the end-to-end efficiency of the transfer, $\alpha_{2}$, and adding them to the harvests of the transmitting node, $\{E_{1,i}\}$. The resulting problem consists of a single energy harvesting link, which can be solved as in \cite{yang2012optimal}. Therefore, in both cases, the power allocation problem reduces to that of a single link. In order to generalize the solution to all cases, we define 
\begin{align}
\alpha_k^*=\max \left( 1,\frac{\alpha_{k}\sigma_j^2 h_j}{\sigma_k^2 h_k} \right),
\end{align}
and find the optimal power allocation policy as the solution to the single user problem with equivalent energy harvests
\begin{align}
\bar{E}_i= \alpha_1^* E_{1,i} + \alpha_2^* E_{2,i}.
\end{align}
The solution is a piecewise constant, non-decreasing sum-power policy, in which the sum-power only changes when all batteries are depleted. A depiction of the optimal sum-power policy is presented in Fig.~\ref{fig_stairs_mac} for $N=4$ time slots, where the staircase represents the cumulative harvested energy and the piecewise linear curve represents the cumulative consumed energy.

\begin{remark}
Optimality of procrastinating policies also extends to channels with more than two transmitters, such as the $K$-user MAC, as shown in \cite{tutuncuoglu2013ita}. In this case, Lemma~\ref{lem_unidirectional} extends to not transferring and receiving energy simultaneously, regardless of the direction \cite[Lemma~1]{tutuncuoglu2013ita}. The procrastination condition extends to the sum of energy transfers arriving to a node \cite[Eqn~12]{tutuncuoglu2013ita}. This allows the iterative directional water-filling algorithms above to be used for such models, by iterating over all transmitters. Due to space restrictions, we refer the reader to \cite{tutuncuoglu2013ita} for details.
\end{remark}

\section{Optimal Policies for Nodes with Finite-sized Batteries}
\label{sect_finite_bat}

We now extend our model, formulation, and solution to nodes with finite-sized batteries. In Lemma~\ref{lem_procrastinating}, it is shown that an optimal procrastinating policy exists if the batteries are infinite-sized. This is justified by always being able to postpone energy transfers which are not consumed within the same time slot, i.e., which do not satisfy (\ref{eqn_lazy1}). In the case of finite-sized batteries, this argument is no longer sufficient, since postponing energy transfers from $T_k$ to $T_j$ may yield a battery overflow at $T_k$ that the original energy transfer policy would have avoided. In this section, we provide a class of policies that procrastinate to the point they can avoid such overflows, and show that they are optimal policies for the EHEC two-way channel with finite-sized batteries.

Consider the finite battery two-way channel model in Section~\ref{sect_model}, i.e., $E_{k}^{max}<\infty$ in (\ref{eqn_model_Bk}). We first reiterate that the optimal policy should not cause any battery overflows. This is an extension of \cite[Lemma 2]{tutuncuoglu2012optimum}, which states that a power policy that yields a battery overflow is suboptimal. In particular, energy overflow in time slot $i$ can be avoided by consuming more energy in time slot $i-1$. This strictly increases the sum-throughput in time slot $i-1$, and does not affect the battery state $S_{k,j}$ for $j=i,\dots,N$. Therefore, without loss of optimality, we restrict our attention to policies that do not cause energy overflows. We use (\ref{eqn_properties_Bk}), while imposing the constraint $S_{k,i} \leq E_k^{max}$ in (\ref{eqn_problem_twc}). The equivalent sum-throughput maximization problem for a TWC with finite-sized batteries is
\begin{subequations}
\label{eqn_problem_twc_finite}
\begin{align}
\label{eqn_problem_twc_finite_a}
\underset{\{p_{k,i},\delta_{k,i}\}}{\max} &  \sum_{i=1}^N C_S^{TWC}(p_{1,i},p_{2,i})\\
\label{eqn_problem_twc_finite_b}
\mbox{s.t.} \quad & E_k^{max} \geq S_{k,i} \geq 0, 	~~ k=1,2, ~ i=1,\dots,N, \\
\label{eqn_problem_twc_finite_c}
& p_{k,i} \geq 0, ~ \delta_{k,i} \geq 0, 			~~ k=1,2, ~ i=1,\dots,N,
\end{align}
\end{subequations}
where $S_{k,i}$ is given by (\ref{eqn_properties_Bk}).

\subsection{Partially Procrastinating Policies}
\label{sub_finite_partial_proc}

We next modify the set of procrastinating policies to prevent energy overflows. We begin by splitting $\delta_{k,i}$ into two components, $\gamma_{k,i} \geq 0$ and $\epsilon_{k,i} \geq 0$, as
\begin{align}
\label{eqn_pro_delta}
	\delta_{k,i}=\gamma_{k,i}+\epsilon_{k,i}.
\end{align}
These components represent the portion of the transferred energy that is consumed immediately, and the excess portion that is stored for future use, respectively. Clearly, power policies defined as $\{p_{k,i},\gamma_{k,i},\epsilon_{k,i}\}$ include all feasible power policies for (\ref{eqn_problem_twc}). Based on these variables, we define {\em partially procrastinating policies}, which are an extension of procrastinating policies in Section~\ref{sub_equivalent_procrastinating}, as follows:
\begin{definition}
\label{def_pro}
	A power policy $\{p_{k,i},\gamma_{k,i},\epsilon_{k,i}\}$ is a partially procrastinating policy if it satisfies
\begin{align}
	\label{eqn_pro_def1}
	p_{k,i} - \alpha_{j} \gamma_{j,i} \geq 0, &\\
	\label{eqn_pro_def2}
	\gamma_{1,i}\gamma_{2,i}=0, &\\
	\label{eqn_pro_def3}
	\epsilon_{k,i} \left( E_k^{max} - S_{k,i} \right) = 0,
\end{align}
for $k,j=1,2$, $j \neq k$, and $i=1,\dots,N$.
\end{definition}

In a {\em partially procrastinating} policy, the condition for procrastination, i.e., (\ref{eqn_lazy1}), is restricted to the immediately consumed component $\gamma_{k,i}$, as seen in (\ref{eqn_pro_def1}). Meanwhile, the excess component $\epsilon_{k,i}$ can only be nonzero if the battery of $T_k$ is full, i.e., $S_{k,i} = E_k^{max}$, as dictated by (\ref{eqn_pro_def3}). This component allows transferring the excess energy that would otherwise be lost due to battery overflows. We next show that there exists at least one optimal policy that is partially procrastinating.

\begin{lemma}
\label{lem_procrastinating_partial}
There exists a partially procrastinating policy $\{p_{k,i},\gamma_{k,i},\epsilon_{k,i}\}$ such that the transferred energy values $\{\delta_{k,i}\}$ calculated from (\ref{eqn_pro_delta}) and the transmit powers $\{p_{k,i}\}$ solve (\ref{eqn_problem_twc_finite}).
\end{lemma}
\begin{IEEEproof}
The proof can be found in Appendix~\ref{app_proof_partial}.
\end{IEEEproof}

\subsection{Finding the Optimal Power Policy}
\label{sub_finite_optimal}

We update the definition of consumed powers in (\ref{eqn_equivalent_consumed}) as
\begin{align}
\label{eqn_finite_consumed}
	\bar{p}_{k,i} = p_{k,i} + \gamma_{k,i} - \alpha_{j} \gamma_{j,i}
\end{align}
for $k,j=1,2$, $j \neq k$, and $i=1,\dots,N$. Substituting in (\ref{eqn_properties_Bk}), this yields the battery state
\begin{align}
\label{eqn_finite_Bk}
	S_{k,i} &=\sum_{n=1}^i \left( E_{k,n}+\alpha_{j} \epsilon_{j,n}-\epsilon_{k,n}-\bar{p}_{k,n} \right).
\end{align}
We next rewrite (\ref{eqn_problem_twc_finite}) in terms of $\bar{p}_{k,i}$, $\gamma_{k,i}$, and $\epsilon_{k,i}$ as
\begin{subequations}
\label{eqn_problem_twc_finite2}
\begin{align}
\label{eqn_problem_twc_finite2_a}
\underset{\{\bar{p}_{k,i},\gamma_{k,i},\epsilon_{k,i}\}} {\max} & \sum_{i=1}^N 
C_S^{TWC} \left( \left[ \bar{p}_{k,i}-\gamma_{k,i}+\alpha_{j} \gamma_{j,i} \right] \right) && &~\\
\label{eqn_problem_twc_finite2_b}
\mbox{s.t.} \quad ~~	& E_k^{max} \geq S_{k,i} \geq 0, 	~~ k=1,2, ~ i=1,\dots,N, \\
\label{eqn_problem_twc_finite2_c}
 				& \hspace{-0.5in} \bar{p}_{k,i} \geq \gamma_{k,i} \geq 0, ~ \gamma_{1,i}\gamma_{2,i}=0, ~~ k=1,2, ~ i=1,\dots,N, \\
\label{eqn_problem_twc_finite2_d}
 				& \hspace{-0.5in} \bar{p}_{k,i} \geq 0, ~ \epsilon_{k,i} \geq 0,  ~~ k=1,2, ~ i=1,\dots,N.
\end{align}
\end{subequations}
In (\ref{eqn_problem_twc_finite2}), we have selectively imposed the partial procrastination conditions (\ref{eqn_pro_def1}) and (\ref{eqn_pro_def2}) without loss of optimality due to Lemma~\ref{lem_procrastinating_partial}. Note that (\ref{eqn_problem_twc_finite2_a}) and (\ref{eqn_problem_twc_finite2_c}) are independent of $\epsilon_{k,i}$, while (\ref{eqn_problem_twc_finite2_b}) and (\ref{eqn_problem_twc_finite2_d}) are independent of $\gamma_{k,i}$. Moreover, the $i$th summation term in (\ref{eqn_problem_twc_finite2_a}) depends on $\bar{p}_{k,i}$, $\gamma_{k,i}$, and $\epsilon_{k,i}$ only, allowing us to decompose (\ref{eqn_problem_twc_finite2}) as
\begin{subequations}
\label{eqn_finite_problemPA}
\begin{align}
\label{eqn_finite_problemPA_a}
	\underset{\{\bar{p}_{k,i},\epsilon_{k,i}\}} {\max} & \sum_{i=1}^N R(\bar{p}_{1,i},\bar{p}_{2,i})\\
\label{eqn_finite_problemPA_b}
	\mbox{s.t.} \quad &  E_k^{max} \geq S_{k,i} \geq 0, ~~ k=1,2, ~ i=1,\dots,N, \\
\label{eqn_finite_problemPA_c}
	& \bar{p}_{k,i} \geq 0, ~ \epsilon_{k,i} \geq 0, ~~ k=1,2, ~ i=1,\dots,N,
\end{align}
\end{subequations}
where $R(\bar{p}_{1,i},\bar{p}_{2,i})$ is the per-slot sum-rate, given by
\begin{subequations}
\label{eqn_finite_problemET}
\begin{align}
\label{eqn_finite_problemET_a}
R(\bar{p}_{1,i},\bar{p}_{2,i}) = \underset{\gamma_{1,i},\gamma_{2,i}} {\max} & ~ C_S^{TWC} \left( \left[ \bar{p}_{k,i}-\gamma_{k,i}+\alpha_{j} \gamma_{j,i} \right] \right)\\
\label{eqn_finite_problemET_b}
\mbox{s.t.} ~~ & ~ \bar{p}_{k,i} \geq \gamma_{k,i} \geq 0, \quad k=1,2,\\
\label{eqn_finite_problemET_c}
 	& ~ \gamma_{1,i}\gamma_{2,i}=0.
\end{align}
\end{subequations}
We remark that (\ref{eqn_finite_problemPA}) and (\ref{eqn_finite_problemET}) are the finite-sized battery extensions of the power allocation and energy transfer problems in (\ref{eqn_problem_PA}) and (\ref{eqn_problem_ET}), respectively. As in Section~\ref{sub_twc_et}, the solution to (\ref{eqn_finite_problemET}) is given by (\ref{eqn_twc_et_solution})-(\ref{eqn_twoway_R}). It remains to solve (\ref{eqn_finite_problemPA}) and identify the optimal $\{\bar{p}_{k,i}\}$ and $\{\epsilon_{k,i}\}$. Observing that (\ref{eqn_finite_problemPA}) is a convex program, we write the KKT optimality conditions 
\begin{align}
\label{eqn_finite_stat_1}
	-\frac{d R(\bar{p}_{1,i},\bar{p}_{2,i})}{d \bar{p}_{k,i}} + \sum_{n=i}^N (\lambda_{k,n} - \beta_{k,n})-\tau_{k,i} &= 0, \\
\label{eqn_finite_stat_2}
	\sum_{n=i}^N (\lambda_{k,n} - \beta_{k,n}) - \alpha_{k} \sum_{n=i}^N (\lambda_{j,n} - \beta_{j,n})-\mu_{k,i} &= 0, \\
	\label{eqn_finite_cs_1}
	\lambda_{k,i} S_{k,i} =0, ~~ \beta_{k,i} (S_{k,i}-E_k^{max}) &=0, \\
\label{eqn_finite_cs_2}
	\tau_{k,i} \bar{p}_{k,i} =0, ~~ \mu_{k,i} \epsilon_{k,i} &=0,
\end{align}
for $k,j=1,2$, $j \neq k$, and $i=1,\dots,N$. Here, $\lambda_{k,i} \geq 0$ and $\beta_{k,i} \geq 0$ are the Lagrange multipliers for the constraints in (\ref{eqn_finite_problemPA_b}), and $\tau_{k,i} \geq 0$ and $\mu_{k,i} \geq 0$ are the Lagrange multipliers for the constraints in (\ref{eqn_finite_problemPA_c}). For the water levels in (\ref{eqn_inf_waterlevel}), (\ref{eqn_finite_stat_1}) yields
\begin{align}
	v_{k,i}= \left( \sum_{n=i}^N (\lambda_{k,n} - \beta_{k,n})	- \tau_{k,i} \right)^{-1}.
\end{align}

For $p_{k,i}>0$, the optimal water levels $v_{k,i}$ may increase only when the battery is empty, $S_{k,i}=0$, and decrease only when the battery is full, $S_{k,i}=E_{k}^{max}$. Omitting (\ref{eqn_finite_stat_2}), this is the finite-sized battery extension of (\ref{eqn_inf_stat_1})-(\ref{eqn_inf_cs_1}), and can be solved by the iterative generalized directional water-filling algorithm where the water flow between time slots is limited to $E_{k}^{max}$ \cite{tutuncuoglu2012sum}. Note that energy cooperation over $\epsilon_{k,i}$ introduces the possibility of energy (and hence water) flow between the two nodes in addition to between two time slots. This flow dimension was first considered in the {\em two dimensional directional water filling algorithm} (2D-DWF) in \cite[Alg. 1]{gurakan2013energycoop} for nodes with infinite batteries. Here, we extend it to the finite-sized battery case, and simplify the algorithm significantly by observing the structure of optimal energy transfers $\{\epsilon_{k,i}\}$.

\begin{lemma}
If the generalized directional water-filling algorithm with finite-sized batteries yields $\bar{p}_{j,i}>0$ for some $i$, then $\epsilon_{k,i}=0$ is optimal, i.e., energy transfer from $T_k$ to $T_j$ with the purpose of storage is not necessary in the optimal policy.
\end{lemma}
\begin{IEEEproof}
Since $\{\bar{p}_{k,i}\}$ is the output of the generalized directional water-filling algorithm, there exists $\lambda_{k,i}$, $\beta_{k,i}$, and $\tau_{k,i}$ that satisfy (\ref{eqn_finite_stat_1}), (\ref{eqn_finite_cs_1}), and (\ref{eqn_finite_cs_2}). Given $\bar{p}_{j,i}>0$, the second sum term in (\ref{eqn_finite_stat_2}) is equal to $v_{j,i}^{-1}$, while the first sum term in (\ref{eqn_finite_stat_2}) is greater than or equal to $v_{k,i}^{-1}$ due to $\tau_{k,i} \geq 0$. By (\ref{eqn_wlevel_1}), water levels always satisfy $v_{k,i} \geq \alpha_{j} v_{j,i} \geq \alpha_k \alpha_j v_{k,i}$, and hence (\ref{eqn_finite_stat_2}) can be satisfied for some $\mu_{k,i}>0$ by choosing $\epsilon_{k,i}=0$.
\end{IEEEproof}

The Lemma implies that it is sufficient to have a nonzero $\epsilon_{k,i}$ only when $\bar{p}_{j,i}=0$. We combine this insight with the condition in (\ref{eqn_pro_def3}), which implies that it is sufficient to have a nonzero $\epsilon_{k,i}$ only when the battery of $T_k$ is full, and propose the {\em 2D-DWF algorithm} {\bf \em with restricted transfers}. In this implementation, we modify the 2D-DWF algorithm by allowing water flow from $T_k$ to $T_j$ only if the battery of $T_k$ is full and $\bar{p}_{j,i}=0$. We allow water flow until (\ref{eqn_finite_stat_2}) is satisfied for $\mu_{k,i}=0$. In accordance with the battery capacity constraint, we also limit the water flow among neighboring time slots to $E_k^{max}$.

\begin{figure}
   \includegraphics[width=\linewidth]{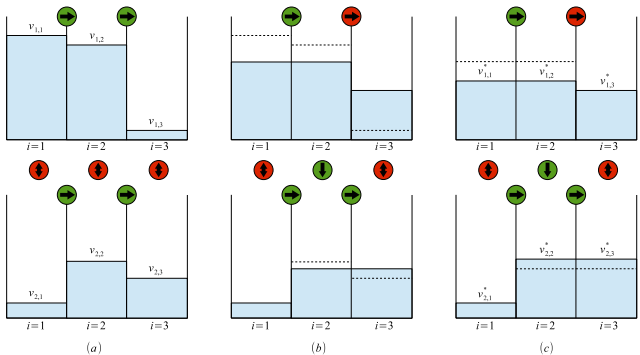}
   \centering
   \caption{Two dimensional directional water-filling with restricted transfers, with (a) initial water levels, (b) water levels after flow within each node, and (c) water levels after flow between the two nodes. The flow from $T_k$ to $T_j$ is not allowed unless the battery of $T_k$ is full, as seen at $i=2$ in (b).}
   \label{fig_finite_example}
   \end{figure}

The 2D-DWF algorithm with restricted transfers is demonstrated in Fig.~\ref{fig_finite_example}. Initially, the entire harvested energy $E_{k,i}$ is allocated to transmission, i.e., $p_{k,i}=E_{k,i}$, and water levels are obtained from (\ref{eqn_wlevel_1}). This state is depicted in Fig.~\ref{fig_finite_example}a. Next, directional water flow is allowed for each user individually and in time only, i.e., flow in the vertical direction is not allowed. The {\em taps} marked with right facing arrows limit water flow to a maximum of $E_k^{max}$ between time slots. The resulting water levels are shown in Fig.~\ref{fig_finite_example}b. Finally, vertical water flow is allowed only in the time slots ending with a full battery, e.g., at $i=2$ from $T_1$ to $T_2$, by turning on the taps marked with vertical arrows. Water flow from $T_1$ to $T_2$ continues until (\ref{eqn_finite_stat_2}) is satisfied, yielding the optimal water levels in Fig.~\ref{fig_finite_example}c. Recall that water flow from $T_k$ to $T_j$, which represents $\epsilon_{k,i}$, is only one component of transferred energy. Energy transfer $\delta_{k,i}$ may be taking place at $i=1,3$ via the immediately consumed component $\gamma_{k,i}$, which are found using (\ref{eqn_twc_et_solution}).

\begin{remark}
The proof of Lemma~\ref{lem_procrastinating_partial} is independent of the objective function of (\ref{eqn_problem_twc_finite}). Hence, the optimality of partially procrastinating policies, shown in Lemma~\ref{lem_procrastinating_partial}, immediately extends to the two-hop and multiple access models in Sections~\ref{sect_twohop} and \ref{sect_mac}. As a result, the 2D-DWF algorithm with restricted transfers can be used for these models as well, provided that the water levels are updated as in (\ref{eqn_thc_wlevel}) for the THC and (\ref{eqn_mac_wlevel}) for the MAC.
\end{remark}

\section{Numerical Results}
\label{sect_numerical}

In our simulations of the three channel models, we consider a transmission period of $N=100$ time slots, a noise spectral density of $N_0=10^{-19}$~W/Hz and a bandwidth of $W=1$~MHz for both nodes. Battery capacities are $E_1^{max}=E_2^{max}=10$~mJ. Unless otherwise stated, the energy arrivals are generated uniformly and independently in $[0,10]$~mJ, the channel power gains are $h_1=h_2=-100$~dB, and the energy transfer efficiency values are $\alpha_1=\alpha_2=0.5$.  
These are typical system parameters similar to those in previous work \cite{gurakan2013energycoop, yang2012optimal, tutuncuoglu2012optimum, tutuncuoglu2012sum, tutuncuoglu2014inefficient, yang2012broadcasting}.
For the purpose of comparison with conventional power allocation policies, we also evaluate the performance of a {\em constant power policy}. Nodes employing the constant power policy attempt transmission with a transmit power equal to their average energy harvesting rate, i.e., $p_{k,i}=\min\{S_{k,i},\mathbb{E}[E_{k,i}]\}$, whenever they are not in an energy outage. To verify that the difference in performance is not solely due to energy cooperation, we additionally allow energy cooperation between nodes employing the constant power policy for their consumed powers, i.e., $\bar{p}_{k,i}=\min\{S_{k,i},\mathbb{E}[E_{k,i}]\}$, while allowing the optimal instantaneous energy transfers given by (\ref{eqn_twc_et_solution}), (\ref{eqn_marc_delta_opt_single}) and below (\ref{eqn_problem_mac_et}).

\subsection{EHEC Two-way Channel}
\label{sub_numerical_twc}

\begin{figure}
   \includegraphics[width=0.9\linewidth]{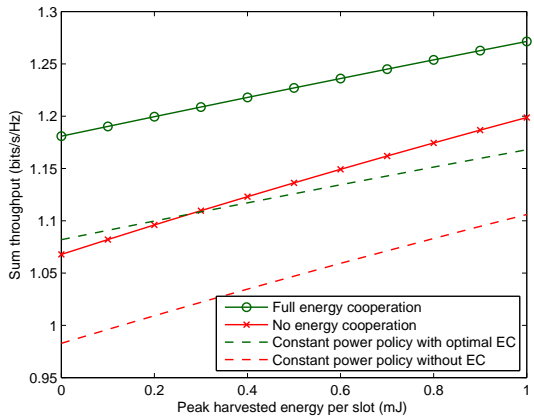}
   \centering
   \caption{Sum-throughput versus peak harvested energy $E_{1,i}$ for the two-way channel with and without energy transfer, compared with the heuristic constant power policy.}
   \label{fig_sim_twc}
   \end{figure}

For the two-way channel, the average sum-throughput values are plotted in Fig.~\ref{fig_sim_twc} for the optimal policy with two-way energy cooperation, optimal policy without energy cooperation, and the two constant power policies. The plots are obtained by varying the peak harvest rate of $T_1$, referred to as $E_h$, in $[0,10]$~mJ, while $E_{1,i}$ is distributed uniformly on $[0,E_h]$. It can be observed that energy cooperation yields a significant increase in performance, particularly since $T_1$ is energy deprived compared to $T_2$. In other evaluations not shown here, a similar insight is observed to hold when one node has a notably worse channel. Constant power policies, on the other hand, perform consistently worse than the respective optimal policies found via generalized directional water-filling.

\begin{figure}
   \includegraphics[width=0.9\linewidth]{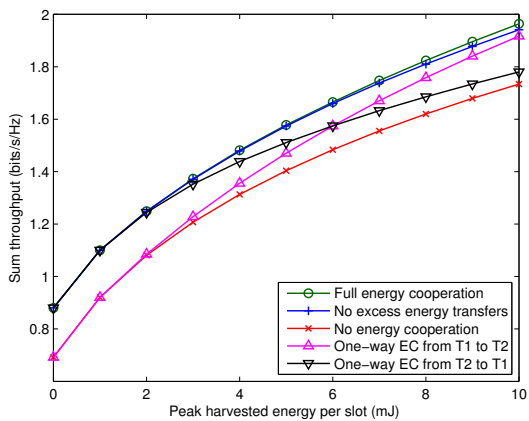}
   \centering
   \caption{Sum-throughput for the two-way channel with one-way energy transfer, and without excess energy transfers $\epsilon_{k,i}$.}
   \label{fig_sim_oneway}
   \end{figure}

In most cases, one node is clearly at a disadvantage in terms of energy, and the direction of optimal energy transfer is usually fixed, i.e., uni-directional energy transfer usually suffices to achieve the maximum throughput with energy cooperation. 
However, cases where bi-directional transfer outperforms uni-directional transfer in either direction are frequently observed. These include cases where the energy budgets and channel parameters of the nodes are comparable, where the energy-deprived node changes within the transmission duration, and where the battery capacity is reached at both nodes at different points in time.
An example is shown in Fig.~\ref{fig_sim_oneway} for the same channel parameters in Fig.~\ref{fig_sim_twc} but with an energy arrival scenario where $T_1$ is energy deprived for one half of the transmission, and $T_2$ is energy deprived for the other. Note that for low or high harvest rates for node $T_1$, the uni-directional energy transfer performs better in different directions, and both directions are essential to achieve the optimal throughput.

\begin{figure}
   \includegraphics[width=0.9\linewidth]{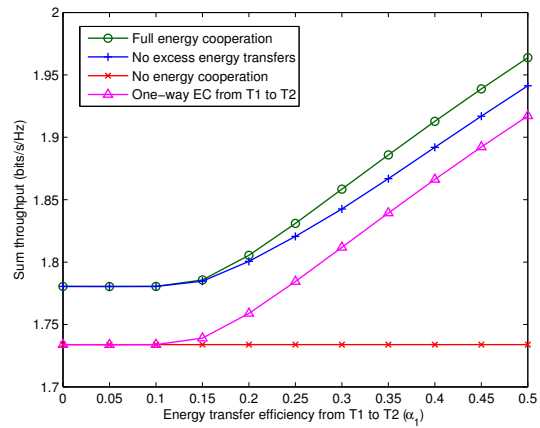}
   \centering
   \caption{Sum-throughput versus transfer efficiency $\alpha_{1}$ for the two-way channel.}
   \label{fig_sim_twc_alpha}
   \end{figure}

Fig.~\ref{fig_sim_oneway} also shows the {\em no excess energy transfers} policy, where the excess energy transfers $\epsilon_{k,i}$ are forced to be zero while instantaneously consumed energy transfers, $\gamma_{k,i}$ are chosen freely. The departure of this policy from the full cooperation case indicates that the excess energy transfers are necessary to find the optimal policy, whereas their impact on throughput is not as significant as the impact of instantaneously consumed energy transfers. This departure can be seen more clearly in Fig.~\ref{fig_sim_twc_alpha}, in which $E_h=10$~mJ is fixed, and the transfer efficiency $\alpha_{1}$ is varied in $[0,\tfrac{1}{2}]$. We also remark that below a certain energy transfer efficiency, energy cooperation in the direction from $T_1$ to $T_2$ is not necessary to achieve the optimal throughput. However, as energy transfer becomes more efficient, i.e., for $\alpha_1>0.1$, the optimal throughput increases for both the uni-directional and bi-directional energy cooperation cases.

\subsection{EHEC Two-hop Channel}
\label{sub_numerical_thc}

\begin{figure}
   \includegraphics[width=0.9\linewidth]{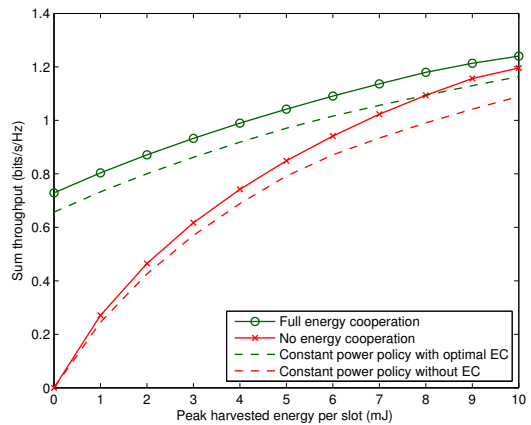}
   \centering
   \caption{Sum-throughput versus peak harvested energy $E_{1,i}$ for the two-hop channel with and without energy transfer, compared with the heuristic constant power policy.}
   \label{fig_sim_thc}
   \end{figure}

We next provide numerical results for the two-hop channel, by varying the peak harvest rate of $T_1$ in $[0,10]$~mJ, in Fig.~\ref{fig_sim_thc}. We observe that in this setup, $T_1$ being energy deprived significantly hinders the performance in the absence of energy cooperation, since both $T_1$ and $T_2$ need to have sufficient energy to transmit in order to have non-zero end-to-end throughput.
Hence, in this case, energy cooperation is observed to be very useful for low $E_h$. Meanwhile, the performance of constant power policies are significantly worse.

\subsection{EHEC Multiple Access Channel}
\label{sub_numerical_mac}

\begin{figure}
   \includegraphics[width=0.9\linewidth]{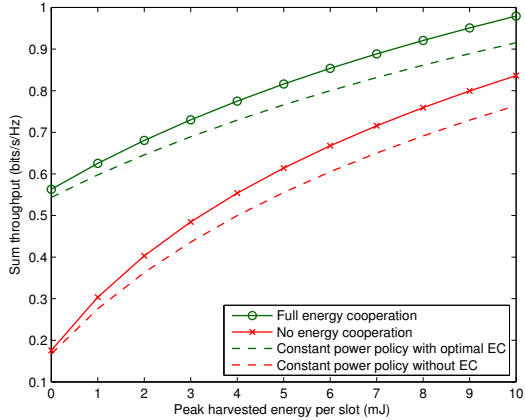}
   \centering
   \caption{Sum-throughput versus peak harvested energy $E_{1,i}$ for the multiple access channel with and without energy transfer, compared with the heuristic constant power policy.}
   \label{fig_sim_mac}
   \end{figure}

Finally, we present numerical results for the two-user multiple access channel, once again varying the peak harvest rate of $T_1$ in $[0,10]$~mJ. The channel gains for users 1 and 2 are $h_1=-100$~dB and $h_2=-110$~dB, respectively. Performance of the optimal policies with and without energy transfer, and constant power policies, are shown in Fig.~\ref{fig_sim_mac}. It can be seen that when $T_1$, which has a better channel to the receiver that that of $T_2$, is energy deprived, energy cooperation from $T_2$ to $T_1$ significantly increases the throughput.

\section{Conclusions}
\label{sect_conclusion}

In this work, we have identified the jointly optimal transmit power and energy cooperation policies of energy harvesting channels for maximizing sum-throughput. This is done by identifying a class of policies that contain an optimal policy and that enable the decomposition of the problem into energy transfer allocation and consumed power allocation problems that can be solved in tandem.

For the two-way channel, the two-hop channel, and the multiple access channel, we have found optimal procrastinating policies using variations of directional water-filling. 
Although the algorithms yielding the optimal policies have the water-filling analogy in common, these models display notable differences that also affect the resulting optimal policies. In particular, in the two-way channel, the two users' communication rates are unaffected by one another, and thus water-filling is performed separately and iteratively for each transmitter. On the other hand, in the two-hop channel, the resulting water-filling algorithm admits a two-fluid interpretation. Finally, in the multiple access channel, the sum-rate is governed by the sum-power, and the problem reduces to a single transmitter counterpart with aggregate arrivals based on the efficiency of energy transfers.
Overall, we have demonstrated how (partially) procrastinating policies simplify the joint optimization problem, both in the infinite and finite battery cases. Simulations have demonstrated that energy cooperation provides a notable increase in sum-throughput, particularly when one node is at a disadvantage in terms of channel power gains or harvesting rate, and the end-to-end throughput depends on both nodes being able to transmit. We have also noted that energy transfer in one direction is not always sufficient, and that bi-directional cooperation is needed to achieve the maximum sum-throughput.

Optimality of procrastinating policies has recently proved to be a useful tool that extends to more involved channel models such as the diamond channel \cite{gurakan2014energy}. 
It would therefore be interesting to extend these results to various other channels. 
Other future directions include considering offline policies for imperfect channel state information or imperfect energy harvest information, online policies inspired by the offline solutions presented in this work, extensions to models with channel fading where channel state information dynamics and availability at various parties can change the underlying rate regions, data buffers, delay constraints, storage inefficiencies, and processing energy cost. Lastly, we note that while we considered a deterministic approach with energy accumulation, transfer, and usage, a queueing treatment of the same remains an interesting future direction.

\appendices

\section{Proof of Lemma~\ref{lem_concavity_of_r}}
\label{app_proof_concavity}

Lemma~\ref{lem_concavity_of_r} follows from the constraints (\ref{eqn_problem_et_b}) being linear, and the function $C_S^{TWC}$ in (\ref{eqn_problem_et_a}) being jointly concave in its arguments. Specifically, let the solution to (\ref{eqn_problem_ET}) for $[\pi_k]$ and $[\tilde{\pi}_k]$ find $[\delta_{k}]$ and $[\tilde{\delta}_{k}]$, respectively. For consumed powers $[a\pi_k + (1-a)\tilde{\pi}_k]$ with $0 \leq a \leq 1$, the energy transfers $[a\delta_{k} + (1-a)\tilde{\delta}_{k}]$ are feasible, and yield 
\begin{align}
R([a\pi_{k} + (1-a)\tilde{\pi}_{k}]) \geq aR([\pi_k])+(1-a)R([\tilde{\pi}_k]),
\end{align}
due to the joint concavity of $C_S^{TWC}(p_1,p_2)$ in $p_1$ and $p_2$.

\section{Proof of Lemma~\ref{lem_procrastinating_partial}}
\label{app_proof_partial}

Let $\{p_{k,i}^*,\delta_{k,i}^*\}$, $k=1,2$, $i=1,\dots,N$, be a solution to (\ref{eqn_problem_twc_finite}). We will construct a partially procrastinating policy $\{p_{k,i}^*,\gamma_{k,i}^*,\epsilon_{k,i}^*\}$ that is feasible. Let $\delta_{k,1}=\delta_{k,1}^*$, $k=1,2$. Starting from $i=1$, we calculate
\begin{align}
\label{eqn_pro_lem1}
	\gamma_{k,i} &=\min \left\{ \delta_{k,i}, \frac{p^*_{j,i}}{\alpha_{k}} \right\}, \\
\label{eqn_pro_lem2}
	\delta_{k,i+1} &=\delta_{k,i+1}^* + \delta^*_{k,i}-\gamma_{k,i},
\end{align}
for $k,j=1,2$, $j \neq k$, and $i=1,\dots,N$. Note that $\{\gamma_{k,i}\}$ in (\ref{eqn_pro_lem1}) satisfy (\ref{eqn_pro_def1}) by definition. Next, for $i=1,\dots,N$, let
\begin{align}
\label{eqn_pro_lem3}
	\gamma_{k,i}^* &=\max \{ 0 , \gamma_{k,i}-\gamma_{j,i} \} , ~~ k,j=1,2,~~j \neq k.
\end{align}
This yields $\{\gamma_{k,i}^*\}$ that satisfy both (\ref{eqn_pro_def1}) and (\ref{eqn_pro_def2}). Let $\epsilon_{k,i}^*=0$ for $k,j=1,2$, $j \neq k$, and $i=1,\dots,N$. Starting from $i=1$, we recalculate $\delta_{k,i}$ from (\ref{eqn_pro_delta}) using $\{\gamma_{k,i}^*,\epsilon_{k,i}^*\}$, and calculate $S_{k,i}$ using the recalculated $\delta_{k,i}$, i.e.,
\begin{align}
\label{eqn_pro_lem4}
S_{k,i}=S_{k,i-1}+E_{k,i}-p_{k,i}-\delta_{k,i}+\alpha_{j}\delta_{j,i}
\end{align}
for $k,j=1,2$ and $j \neq k$, while updating the optimal stored component $\epsilon_{k,i}^*$ as
\begin{align}
\label{eqn_pro_lem5}
	\epsilon_{k,i}^* &=\max \{ 0 , S_{k,i}-E_k^{max} \} .
\end{align}
Note that this immediately satisfies (\ref{eqn_pro_def3}) for all $k,j=1,2$ and $i=1,\dots,N$.

The process outlined above postpones energy transfers that are not immediately needed via (\ref{eqn_pro_lem1})-(\ref{eqn_pro_lem2}), eliminates cases of simultaneous bi-directional energy transfer via (\ref{eqn_pro_lem3}), and transfers excess energy that is overflowing via (\ref{eqn_pro_lem5}). The resulting policy, $\{p_{k,i}^*,\gamma_{k,i}^*,\epsilon_{k,i}^*\}$, is a partially procrastinating policy. Given that the original policy $\{p_{k,i}^*,\delta_{k,i}^*\}$ is feasible, $\{p_{k,i}^*,\gamma_{k,i}^*,\epsilon_{k,i}^*\}$ is also feasible by construction. This policy is also optimal since the objective of (\ref{eqn_problem_twc_finite}) depends only on the transmit powers $\{p_{k,i}\}$, and the transmit powers $\{p_{k,i}^*\}$ are equal in both policies.


\begin{thebibliography}{10}
\providecommand{\url}[1]{#1}
\csname url@samestyle\endcsname
\providecommand{\newblock}{\relax}
\providecommand{\bibinfo}[2]{#2}
\providecommand{\BIBentrySTDinterwordspacing}{\spaceskip=0pt\relax}
\providecommand{\BIBentryALTinterwordstretchfactor}{4}
\providecommand{\BIBentryALTinterwordspacing}{\spaceskip=\fontdimen2\font plus
\BIBentryALTinterwordstretchfactor\fontdimen3\font minus
  \fontdimen4\font\relax}
\providecommand{\BIBforeignlanguage}[2]{{%
\expandafter\ifx\csname l@#1\endcsname\relax
\typeout{** WARNING: IEEEtran.bst: No hyphenation pattern has been}%
\typeout{** loaded for the language `#1'. Using the pattern for}%
\typeout{** the default language instead.}%
\else
\language=\csname l@#1\endcsname
\fi
#2}}
\providecommand{\BIBdecl}{\relax}
\BIBdecl

\bibitem{kansal2007power}
A.~Kansal, J.~Hsu, S.~Zahedi, and M.~B. Srivastava, ``Power management in
  energy harvesting sensor networks,'' \emph{ACM Trans. on Embedded Computing
  Systems}, vol.~6, no.~4, pp. 32--69, Sep. 2007.

\bibitem{ulukus2015review}
S.~Ulukus, A.~Yener, E.~Erkip, O.~Simeone, M.~Zorzi, P.~Grover, and K.~Huang,
  ``Energy harvesting wireless communications: {A} review of recent advances,''
  \emph{IEEE Jour. on Selected Areas in Comm.}, vol.~33, no.~3, pp. 360--381,
  Mar. 2015.

\bibitem{xiao2015survey}
L.~Xiao, P.~Wang, D.~Niyato, D.~Kim, and Z.~Han, ``Wireless networks with {RF}
  energy harvesting: {A} contemporary survey,'' \emph{IEEE Comm. Surveys
  Tutorials}, vol.~17, no.~2, pp. 757--789, May 2015.

\bibitem{kurs2007wireless}
A.~Kurs, A.~Karalis, R.~Moffatt, J.~D. Joannopoulos, P.~Fisher, and
  M.~Solja{\v{c}}i{\'c}, ``Wireless power transfer via strongly coupled
  magnetic resonances,'' \emph{Science}, vol. 317, no. 5834, pp. 83--86, Jul.
  2007.

\bibitem{gurakan2012energy}
B.~Gurakan, O.~Ozel, J.~Yang, and S.~Ulukus, ``Energy cooperation in energy
  harvesting wireless communications,'' in \emph{Proc. IEEE International
  Symposium on Information Theory, ISIT}, Jul. 2012.

\bibitem{gurakan2012two}
------, ``Two-way and multiple-access energy harvesting systems with energy
  cooperation,'' in \emph{Proc. Asilomar Conference on Signals, Systems, and
  Computers}, Nov. 2012.

\bibitem{gurakan2013energycoop}
------, ``Energy cooperation in energy harvesting communications,'' \emph{IEEE
  Trans. on Comm.}, vol.~61, no.~12, pp. 4884--4898, Dec. 2013.

\bibitem{yang2012optimal}
J.~Yang and S.~Ulukus, ``Optimal packet scheduling in an energy harvesting
  communication system,'' \emph{IEEE Trans. on Comm.}, vol.~60, no.~1, pp.
  220--230, Jan. 2012.

\bibitem{tutuncuoglu2012optimum}
K.~Tutuncuoglu and A.~Yener, ``Optimum transmission policies for battery
  limited energy harvesting nodes,'' \emph{IEEE Trans. on Wireless Comm.},
  vol.~11, no.~3, pp. 1180--1189, Mar. 2012.

\bibitem{ozel2011fading}
O.~Ozel, K.~Tutuncuoglu, J.~Yang, S.~Ulukus, and A.~Yener, ``Transmission with
  energy harvesting nodes in fading wireless channels: Optimal policies,''
  \emph{IEEE Jour. on Selected Areas in Comm.}, vol.~29, no.~8, pp. 1732--1743,
  Sep. 2011.

\bibitem{yang2012mac}
J.~Yang and S.~Ulukus, ``Optimal packet scheduling in a multiple access channel
  with energy harvesting transmitters,'' \emph{Jour. of Comm. and Networks,
  Special Issue on Energy Harvesting in Wireless Networks}, vol.~14, no.~2, pp.
  140--150, Apr. 2012.

\bibitem{tutuncuoglu2012sum}
K.~Tutuncuoglu and A.~Yener, ``Sum-rate optimal power policies for energy
  harvesting transmitters in an interference channel,'' \emph{Jour. of Comm.
  and Networks, Special Issue on Energy Harvesting in Wireless Networks},
  vol.~14, no.~2, pp. 151--161, Apr. 2012.

\bibitem{huang2013throughput}
C.~Huang, R.~Zhang, and S.~Cui, ``Throughput maximization for the {Gaussian}
  relay channel with energy harvesting constraints,'' \emph{IEEE Jour. on
  Selected Areas in Comm.}, vol.~31, no.~8, pp. 1469--1479, Aug. 2013.

\bibitem{gunduz2011two}
D.~Gunduz and B.~Devillers, ``Two-hop communication with energy harvesting,''
  in \emph{Proc. IEEE International Workshop on Computational Advances in
  Multi-Sensor Adaptive Processing, CAMSAP}, Dec. 2011.

\bibitem{orhan2013throughput}
O.~Orhan and E.~Erkip, ``Throughput maximization for energy harvesting two-hop
  networks,'' in \emph{Proc. IEEE International Symposium on Information
  Theory, ISIT}, Jul. 2013.

\bibitem{ahmed2012power}
I.~Ahmed, A.~Ikhlef, R.~Schober, and R.~K. Mallik, ``Power allocation for
  conventional and buffer-aided link adaptive relaying systems with energy
  harvesting nodes,'' \emph{IEEE Trans. on Wireless Comm.}, vol.~13, no.~3, pp.
  1182--1195, Mar. 2014.

\bibitem{luo2012optimal2}
Y.~Luo, J.~Zhang, and K.~B. Letaief, ``Optimal scheduling and power allocation
  for two-hop energy harvesting communication systems,'' \emph{IEEE Trans. on
  Wireless Comm.}, vol.~12, no.~9, pp. 4729--4741, Sep. 2013.

\bibitem{varan2013energy}
B.~Varan and A.~Yener, ``The energy harvesting two-way decode-and-forward relay
  channel with stochastic data arrivals,'' in \emph{Proc. IEEE GlobalSIP
  Symposium on Energy Harvesting and Green Wireless Comm.}, Dec. 2013.

\bibitem{tutuncuoglu2012ita}
K.~Tutuncuoglu and A.~Yener, ``Communicating with energy harvesting
  transmitters and receivers,'' in \emph{Proc. Information Theory and
  Applications Workshop, ITA}, Feb. 2012.

\bibitem{mahdavi2013energy}
H.~Mahdavi-Doost and R.~D. Yates, ``Energy harvesting receivers: Finite battery
  capacity,'' in \emph{Proc. IEEE International Symposium on Information
  Theory, ISIT}, Jul. 2013.

\bibitem{niyato2007sleep}
D.~Niyato, E.~Hossain, and A.~Fallahi, ``Sleep and wakeup strategies in
  solar-powered wireless sensor/mesh networks: Performance analysis and
  optimization,'' \emph{IEEE Trans. on Mobile Computing}, vol.~6, no.~2, pp.
  221--236, Feb. 2007.

\bibitem{want2006introduction}
R.~Want, ``An introduction to {RFID} technology,'' \emph{IEEE Pervasive
  Computing}, vol.~5, no.~1, pp. 25--33, Jan. 2006.

\bibitem{grover2010shannon}
P.~Grover and A.~Sahai, ``{S}hannon meets {T}esla: {W}ireless information and
  power transfer,'' in \emph{Proc. IEEE International Symposium on Information
  Theory, ISIT}, Jun. 2010.

\bibitem{popovski2012interactive}
P.~Popovski, A.~M. Fouladgar, and O.~Simeone, ``Interactive joint transfer of
  energy and information,'' \emph{IEEE Trans. on Comm.}, vol.~61, no.~5, pp.
  2086--2097, May 2013.

\bibitem{ng2013wireless}
D.~W.~K. Ng, E.~S. Lo, and R.~Schober, ``Energy-efficient resource allocation
  in multiuser {OFDM} systems with wireless information and power transfer,''
  in \emph{Proc. IEEE Wireless Comm. and Networking Conference, WCNC}, Apr.
  2013.

\bibitem{michalopoulos2014simultaneous}
D.~S. Michalopoulos, H.~A. Suraweera, and R.~Schober, ``Simultaneous
  information transmission and wireless energy transfer via selecting one out
  of two relays,'' in \emph{Proc. 6th International Symposium on
  Communications, Control and Signal Processing, ISCCSP}, Apr. 2014.

\bibitem{michalopoulos2013relay}
------, ``Relay selection for simultaneous information transmission and
  wireless energy transfer: {A} tradeoff perspective,'' \emph{IEEE Jour. on
  Selected Areas in Comm.}, vol.~33, Sep. 2015, [Online]
  http://arxiv.org/abs/1303.1647.

\bibitem{chen2013energy}
X.~Chen, X.~Wang, and X.~Chen, ``Energy-efficient optimization for wireless
  information and power transfer in large-scale {MIMO} systems employing energy
  beamforming,'' \emph{IEEE Wireless Comm. Letters}, vol.~2, no.~6, pp.
  667--670, Dec. 2013.

\bibitem{shannon1961two}
C.~E. Shannon, ``Two-way communication channels,'' in \emph{Proc. Berkeley
  Symposium on Mathematical Statistics and Probability}, 1961.

\bibitem{bertsekas1999nonlinear}
D.~P. Bertsekas, \emph{Nonlinear programming}.\hskip 1em plus 0.5em minus
  0.4em\relax Athena Scientific, Belmont, MA, 1999.

\bibitem{goldsmith1997capacity}
A.~J. Goldsmith and P.~Varaiya, ``Capacity of fading channels with channel side
  information,'' \emph{IEEE Trans. on Information Theory}, vol.~43, no.~6, pp.
  1986--1992, Nov. 1997.

\bibitem{tutuncuoglu2013ita}
K.~Tutuncuoglu and A.~Yener, ``Multiple access and two-way channels with energy
  harvesting and bidirectional energy cooperation,'' in \emph{Proc. Information
  Theory and Applications Workshop, ITA}, Feb. 2013.

\bibitem{tutuncuoglu2014inefficient}
K.~Tutuncuoglu, A.~Yener, and S.~Ulukus, ``Optimum policies for an energy
  harvesting transmitter under energy storage losses,'' \emph{IEEE Jour. on
  Selected Areas in Comm.: Wireless Comm. Powered by Energy Harvesting and
  Wireless Energy Transfer}, vol.~33, no.~3, pp. 467--481, Mar. 2015.

\bibitem{yang2012broadcasting}
J.~Yang, O.~Ozel, and S.~Ulukus, ``Broadcasting with an energy harvesting
  rechargeable transmitter,'' \emph{IEEE Trans. on Wireless Comm.}, vol.~11,
  no.~2, pp. 571--583, Feb. 2012.

\bibitem{gurakan2014energy}
B.~Gurakan and S.~Ulukus, ``Energy harvesting diamond channel with energy
  cooperation,'' in \emph{Proc. IEEE International Symposium on Information
  Theory, ISIT}, Jun. 2014.

\end{thebibliography}
\end{document}